\documentclass[float,epsfig,usenatbib]{mn2e}

\usepackage{graphicx}
\usepackage{appendix}
\usepackage{psfrag}
\usepackage{amsmath,amssymb}
\usepackage{threeparttable}
\usepackage{float}
\usepackage{subfigure}
\usepackage{afterpage}

\title[The GALEX Arecibo SDSS Survey. I.]{The GALEX Arecibo SDSS
  Survey. I. Gas Fraction Scaling Relations of Massive Galaxies and First Data Release}
\author[B. Catinella et al.]
{Barbara Catinella$^{1}$\thanks{bcatinel@mpa-garching.mpg.de},
  David Schiminovich$^{2}$, Guinevere Kauffmann$^{1}$, Silvia Fabello$^{1}$, 
  \newauthor Jing Wang$^{1,3}$, Cameron Hummels$^{2}$, Jenna Lemonias$^{2}$, Sean M. Moran$^{4}$, Ronin Wu$^{5}$,
  \newauthor Riccardo Giovanelli$^{6}$, Martha P. Haynes$^{6}$, Timothy M. Heckman$^{4}$, Antara R. 
  \newauthor Basu-Zych$^{7}$, Michael R. Blanton$^{5}$, Jarle Brinchmann$^{8,9}$, Tam{\'a}s Budav{\'a}ri$^{4}$, Thiago 
  \newauthor Gon{\c c}alves$^{10}$, Benjamin D. Johnson$^{11}$, Robert C. Kennicutt$^{11,12}$, 
  \newauthor Barry F. Madore$^{13}$, Christopher D. Martin$^{10}$, Michael R. Rich$^{14}$, Linda J. Tacconi$^{15}$, 
  \newauthor David A. Thilker$^{4}$, Vivienne Wild$^{16}$, and Ted K. Wyder$^{10}$\\
$^{1}$Max-Planck Institut f\"{u}r Astrophysik, D-85741 Garching, Germany\\
$^{2}$Department of Astronomy, Columbia University, New York, NY 10027, USA\\
$^{3}$Center for Astrophysics, University of Science and Technology of China, 230026 Hefei, China\\
$^{4}$Department of Physics and Astronomy, The Johns Hopkins University, Baltimore, MD 21218, USA\\
$^{5}$Department of Physics, New York University, New York, NY 10003 USA\\
$^{6}$Center for Radiophysics and Space Research, Cornell University, Ithaca, NY 14853, USA\\
$^{7}$NASA Goddard Space Flight Center, Laboratory for X-ray Astrophysics, Greenbelt, MD 20771, USA\\
$^{8}$Leiden Observatory, Leiden University, 2300 RA, Leiden, The Netherlands\\
$^{9}$Centro de Astrof\'{\i}sica, Universidade do Porto, 4150-762 Porto, Portugal\\
$^{10}$California Institute of Technology, Pasadena, CA 91125, USA\\
$^{11}$Institute of Astronomy, Cambridge CB3 0HA, UK\\
$^{12}$Steward Observatory, University of Arizona, Tucson, AZ 85721, USA\\
$^{13}$Observatories of the Carnegie Institution of Washington, Pasadena, CA 91101, USA\\
$^{14}$Department of Physics and Astronomy, University of California, Los Angeles, CA 90095, USA\\
$^{15}$Max Planck Institut f\"{u}r extraterrestrische Physik, D-85741 Garching, Germany\\
$^{16}$Institut d'Astrophysique de Paris, 75014 Paris, France      
}

\date{}

\begin{document}

\def\deg{$^{\circ} $}
\newcommand{\eg}{{\it e.g.}}
\newcommand{\ie}{{\it i.e.}}
\newcommand{\minusone}{$^{-1}$}
\newcommand{\kms}{km~s$^{-1}$}
\newcommand{\kmsm}{km~s$^{-1}$~Mpc$^{-1}$}
\newcommand{\Ha}{$\rm H\alpha$}
\newcommand{\Hb}{$\rm H\beta$}
\newcommand{\hi}{{H{\sc i}}}
\newcommand{\hii}{{H{\sc ii}}}
\newcommand{\nii}{\ion{N}{2}}
\newcommand{\rband}{{\em r}-band}
\newcommand{\iband}{{\em I}-band}
\newcommand{\zband}{{\em z}-band}
\newcommand{\rd}{$r_{\rm d}$}
\newcommand{\whi}{$W_{50}$}
\newcommand{\x}{$\times$}
\newcommand{\about}{$\sim$}
\newcommand{\Msun}{M$_\odot$}
\newcommand{\Lsun}{L$_\odot$}
\newcommand{\Mhi}{$M_{\rm HI}$}
\newcommand{\Mst}{$M_\star$}
\newcommand{\must}{$\mu_\star$}
\newcommand{\nuvr}{NUV$-r$}
\newcommand{\Rinz}{$R_{50,z}$}
\newcommand{\tmax}{$T_{\rm max}$}
\newcommand{\faa}{$f_{\rm AA}$}
\newcommand{\fhiar}{$f_{\rm S05}$}
\newcommand{\naa}{$N_{\rm AA}$}
\newcommand{\nhiar}{$N_{\rm S05}$}
\newcommand{\naab}{$N'_{\rm AA}$}
\newcommand{\nhiarb}{$N'_{\rm S05}$}
\newcommand{\ngass}{$N_{\rm G}$}
\newcommand{\ngassnorich}{$N_{\rm G,<AA}$}

\maketitle

\label{firstpage}

\begin{abstract}

We introduce the GALEX Arecibo SDSS Survey (GASS), an on-going large
program that is gathering high quality \hi-line spectra using the
Arecibo radio telescope for an unbiased sample of \about 1000 galaxies
with stellar masses greater than $10^{10}$ \Msun\ and redshifts
$0.025 < z < 0.05$, selected from the SDSS spectroscopic and GALEX imaging surveys. 
The galaxies are observed until detected or until a low gas mass
fraction limit (1.5$-$5\%) is reached. This paper presents the first
Data Release, DR1, consisting of  \about 20\% of the final GASS sample.
We use this data set to explore the main scaling relations of
\hi\ gas fraction with galaxy structure and NUV-r colour.
A large fraction (\about 60\%) of the galaxies in our sample are
detected in \hi. Even at stellar masses above $10^{11}$ \Msun, the
detected fraction does not fall below \about 40\%.  
We find that the atomic gas fraction  \Mhi/\Mst\ decreases 
strongly with stellar mass, stellar surface mass density and
\nuvr\ colour, but is only weakly correlated with galaxy bulge-to-disk ratio
(as measured by the concentration index of the \rband\ light).
We also find that the fraction of galaxies with significant (more than a few percent) 
\hi\ decreases sharply above a characteristic stellar surface mass density of  
$10^{8.5}$ \Msun~kpc$^{-2}$.  The fraction of gas-rich galaxies
decreases much more smoothly with stellar mass.   
One of the key goals of the GASS survey is to identify and
quantify the incidence of galaxies that are {\em transitioning} 
between the blue, star-forming cloud and the red sequence of
passively-evolving galaxies. Likely transition candidates can be identified as
outliers from the mean scaling relations between \Mhi/\Mst\ and other galaxy properties. 
We have fit a plane to the 2-dimensional relation between \hi\ mass fraction,
stellar surface mass density, and \nuvr\ colour. 
Interesting outliers from this plane include gas-rich red sequence
galaxies that may be in the process of regrowing their disks, 
as well as blue, but gas-poor spirals.
\end{abstract}


\section{Introduction}\label{s_intro}
 
There are good reasons to investigate how and why the cold gas content
of a massive galaxy varies with stellar mass and other physical properties relating to
its  growth history.  While the distinction between red, old
ellipticals and blue, star-forming spirals has been known for a long
time, recent work based on the Sloan Digital Sky Survey
\citep[SDSS;][]{sdss} has shown that galaxies appear
to divide into two distinct ``families'' at a stellar mass
\Mst \about 3 \x $10^{10}$ \Msun\ \citep{str01,kau03b,bal04}.
Lower mass galaxies typically have young stellar populations, low
surface mass densities and the low concentrations characteristic of
disks. On the other hand, galaxies with old stellar populations, high
surface mass densities and the high concentrations typical of bulges
tend to have higher mass.
New theoretical work has led to a diverse set of possible mechanisms
to explain this characteristic mass scale where galaxies transition
from young to old {\citep{keres05,dekel06,hopkins08}, with nearly all
operating via quenching or regulation of the gas supply
\citep[\eg,][]{martin07}.  Observations
of the cold \hi\ gas component -- the  source of the material that will
eventually form stars -- in galaxies across the transition mass, will
provide an important new test of those models.

Initial clues can come from the study of the \hi\ scaling relations of
massive galaxies. Early seminal work (\eg, \citealt{hg84,ktc85,rob91};
see \citealt{rh94} for a review) took the first step towards answering the
question of how the \hi\ properties of galaxies vary as a function of
morphological type, environment and other physical properties.
Although massive ``transition'' galaxies are found in these and more
recent samples (\citealt{pat03}; \citealt{s05}, hereafter S05;
\citealt{bothwell09}), the selection criteria make it difficult to
identify these in a robust way. One needs to quantify the 
{\it distribution} of gas fraction as a function of
\Mst, colour, star formation rate (SFR) and other galaxy
properties in order to understand which galaxies are more gas rich or
gas poor than the mean. 

It has become common practice to use ``photometric
gas-fractions'' \citep{bell03,kan04,zhang09}, which exploit the
well-known connection between SFR and gas content
\citep{schmidt59,ken98}, as a substitute for real gas measurements.
However, these derived average relations cannot be used to study if
and how the gas content relates to other properties and physical
conditions in the galaxies.

\hi\ studies of transition objects require large and uniform samples 
spanning a wide range in gas fraction, stellar mass and other
galaxy properties (\eg, structural parameters and star formation).
Although blind surveys offer the required uniformity, \hi\ studies of transition galaxies
are currently not possible because the depths
reached by existing wide-area blind \hi\ surveys are very shallow compared to surveys such as the
SDSS. The \hi\ Parkes All-Sky Survey \citep{hipass,m04},
covered $\sim 30000$ deg$^2$ and produced a final
catalog of around 5000 \hi\ detections with a median redshift of 2800 \kms. This should be
contrasted with the main SDSS spectroscopic survey, which covers around 7000 deg$^2$ and
contains more than half a million galaxies with a median redshift of 30,000 \kms.
The recently initiated Arecibo Legacy Fast ALFA survey \citep[ALFALFA;][]{alfalfa}
is mapping 7000 deg$^2$ to considerably deeper limits.
With a median redshift of \about 9100 \kms, ALFALFA for the first
time samples the \hi\ population over a cosmologically fair volume,
and is expected to detect \about 30,000 extragalactic \hi-line sources out to
redshifts of $z\sim0.06$. Even so, the galaxies in the transition
regime detected with ALFALFA will be predominantly gas-rich.

In this paper we describe the first results and data release from the
GALEX Arecibo SDSS Survey\footnote{
{\em http://www.mpa-garching.mpg.de/GASS/}
}
(GASS), a new \hi\ survey specifically
designed to obtain \hi\ measurements of \about 1000
galaxies in the local universe ($0.025<z<0.05$) with stellar masses 
$M_\star > 10^{10}$ \Msun.  As we discuss below, we observe these massive
galaxies down to a low gas mass fraction limit (1.5$-$5\%), in order to
study the physical mechanisms that shape the stellar mass function,
regulate gas accretion and quench further galaxy growth by
conversion of gas into stars.   We expect that GASS will provide a
rich, homogeneous data set of structural and physical parameters (\eg,
luminosity, stellar mass, size, surface brightness, gas-phase and
stellar metallicities, AGN content, velocity dispersion), star
formation rates and gas properties.
Analysis of this unique sample should allow us, for the first time,
to investigate how the cold gas responds to a variety of different
physical conditions in the galaxy and obtain new insights on
the physical processes responsible for the transition between blue,
star-forming spirals and red, passively-evolving ellipticals.

In a companion paper (Schiminovich et al., in preparation;
hereafter Paper~II), we derive volume-averaged
quantities for our GASS sample to determine the relative fraction
of \hi\ associated with massive galaxies in the local universe, and
compare with the SFR density to explore how the gas consumption timescale
varies across the galaxy population.  In both papers we also discuss
how we expect to refine and improve our analyses using the full GASS data set.

In this first paper we describe GASS survey design and
selection criteria (\S~\ref{s_design} and \S~\ref{s_sample}). Arecibo
observations and data processing are discussed in \S~\ref{s_obs}. 
We provide catalogs of SDSS/GALEX
parameters and \hi-line spectroscopy measurements for the 176 galaxies
in this first Data Release (DR1) in \S~\ref{s_sdss} and
\S~\ref{s_hi}, respectively. Our results are presented in \S~\ref{s_res}. 
We characterize the properties of the DR1 data set in
\S~\ref{s_dr1}. In order to obtain a sample that is unbiased in terms of
\hi\ properties, we need to correct for the fact
that we do not re-observe objects already detected by ALFALFA or
galaxies found in the Cornell \hi\ archive (S05). We construct such a 
{\em representative} sample in \S~\ref{s_repr}, and we use it 
to quantify average gas fraction scaling relations as a function of
other galaxy parameters in \S~\ref{s_gf} and \S~\ref{s_plane}. 
Our findings are summarized and further discussed in \S~\ref{s_disc}.

All the distance-dependent quantities in this work are computed
assuming $\Omega=0.3$, $\Lambda=0.7$ and $H_0 = 70$ \kmsm. 
AB magnitudes are used throughout the paper.

\section{Survey Design}\label{s_design}

GASS is designed to efficiently measure the \hi\ content of 
an unbiased sample of \about 1000 massive galaxies, for which SDSS
spectroscopy and GALEX \citep{galex} imaging are also available. As described below,
the targets are selected only by redshift and stellar mass, and
observed with the Arecibo radio telescope until detected or until a gas
fraction limit of $1.5-5$\% is reached (\ie, a gas fraction limit an
order of magnitude lower than in objects of similar stellar mass
detected by ALFALFA at the same redshifts). We describe below our survey
requirements and sample selection methodology.

{\bf Survey footprint.} All the GASS targets are located within the
intersection of the footprints of the SDSS primary spectroscopic
survey, the projected GALEX Medium Imaging Survey (MIS) and ALFALFA. The SDSS
primary spectroscopic sample targets all galaxies with $r < 17.77$
with high completeness ($>80$\% for $r > 14.5$). 
The GALEX MIS reaches limiting NUV magnitude
\about 23, which allows us to probe the full range of colours (and
derived SFRs) of the normal galaxy population. Existing ALFALFA coverage
increases our survey efficiency by allowing us to remove from
the GASS target list any objects already detected by ALFALFA --- this
amounts to an estimated 20\% of the galaxies meeting our selection
criteria. However, this does not correspond to a 20\% gain in
observing time, because the objects that we skip are those that
would be detected with the shortest integrations.
We also do not re-observe galaxies with detections in the Cornell
\hi\ digital archive of targeted observations (S05), 
a homogeneous compilation of \hi\ parameters for \about 9000
optically-selected galaxies, mostly selected for Tully-Fisher applications.

{\bf Stellar mass range ($10 < {\rm Log} M_\star/M_\odot < 11.5$).} We
target a stellar mass range that straddles the ``transition
mass'' at \about $3 \times 10^{10}$ \Msun.

{\bf Redshift range ($0.025 < z < 0.05$).} An \hi\ survey of massive
galaxies selected from SDSS is ideally performed at redshifts above $z
> 0.025$. At magnitudes brighter than $r \sim 13$ (corresponding to a
transition mass galaxy at $z=0.025$), the spectroscopic targeting
completeness in the SDSS falls below 50\%.  Additionally, a single
pointing on galaxies at $z<0.025$ may occasionally underestimate the
\hi\ flux, if their \hi\ disks are extended in comparison with the
Arecibo beam. The upper end of our redshift interval is set by
practical sensitivity limits as well as a desire to remain within the
velocity range covered by ALFALFA ($0 < z < 0.06$). 
We further restricted the range to $z < 0.05$ in order to avoid the gap in
ALFALFA velocity coverage caused by radio frequency interference (RFI) 
at 1350 MHz from the the Federal Aviation Administration (FAA) radar in San Juan.

{\bf Gas mass fraction/gas mass limit.} A crucial goal of GASS is to
identify galaxies that show signs of recent accretion and/or
quenching. We wish to obtain accurate gas mass measurements for
transition galaxies, which have had a small, but significant amount of
recent star formation ($1-5$\% of their total mass). This translates
into a requirement that we observe the sample to an equivalent gas
mass fraction (defined as \Mhi/\Mst\ 
in this work) limit. Practically, we have set a limit of 
$M_{\rm HI}/M_\star > 0.015$ for galaxies with \Mst $> 10^{10.5}$ \Msun, and a
constant gas mass limit \Mhi $=10^{8.7}$ \Msun\ for galaxies with
smaller stellar masses. This corresponds to a gas fraction limit $0.015-0.05$ 
for the whole sample. This allows us to detect galaxies with gas
fractions significantly below those of the \hi-rich ALFALFA
detections at the same redshifts, and find early-type transition
galaxies harboring significant reservoirs of gas. We do not try
to detect inconsequential amounts of gas ($M_{\rm HI}/M_\star < 0.01$)
typical of the most gas-poor early-types.

Based on the \hi\ mass limit assigned to each galaxy (\ie, 
\Mhi $= 10^{8.7}$ \Msun\ or 0.015 \Mst, whichever is larger), we have
computed the observing time, \tmax, required to reach that value with our
observing mode and instrumental setup (see \S~\ref{s_sdss}).

\section{Sample Selection}\label{s_sample}

Since the ALFALFA and GALEX surveys are on-going, we have defined a
GASS {\em parent sample}, based on SDSS DR6 \citep{sdssdr6} and the maximal ALFALFA
footprint, from which the targets for Arecibo observations are
extracted. The parent sample includes 12006 galaxies that
satisfy our stellar mass and redshift selection criteria (see
Fig.~\ref{ps_skyd}); of these, \about 10000 have UV photometry from
either the GALEX All-sky Imaging Survey (AIS; \about 100 s exposure, 
FUV$_{lim}$, NUV$_{lim} < 21~m_{\rm AB}$) or the MIS 
(\about 1500 s exposure, FUV$_{lim}$, NUV$_{lim} < 23~m_{\rm AB}$).
The final GASS sample will include \about 1000 galaxies,
chosen by randomly selecting a subset which balances the distribution
across stellar mass and which maximizes existing GALEX exposure
time. 
In practice, we have extracted a subset of few hundred targets distributed
across the whole GASS footprint, which includes galaxies that already
have MIS or at least AIS data, and we have given highest priority to
those in the sky regions already catalogued by ALFALFA. We have also
given some priority to objects with stellar mass greater than
10$^{10.5}$ \Msun, but the final survey sample will have similar
numbers of galaxies in each stellar mass bin.

\begin{figure*}
\includegraphics[width=14cm]{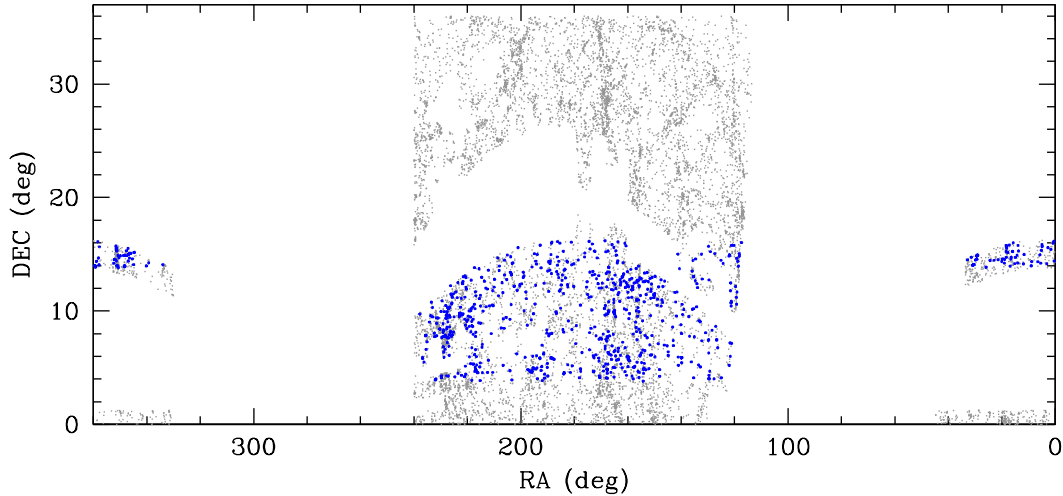}
\caption{Sky footprint of the GASS survey. Gray points represent galaxies in the
{\em parent sample}, the super-set of \about 12000 objects out of
which targets for Arecibo observations are extracted. Blue points
indicate the subset of galaxies detected by the ALFALFA survey to date.}
\label{ps_skyd}
\end{figure*}

\subsection{Overlap with ALFALFA and Cornell Digital \hi\ archive}\label{s_overlap}

As mentioned in \S~\ref{s_design}, we do not re-observe galaxies with
good \hi\ measurements already available from either ALFALFA or the S05
digital archive. The overlap between the GASS parent sample and the
S05 compilation of detected galaxies (their Table 3) is 430 objects.
ALFALFA has released five catalogs to date, three of which are in sky
regions with SDSS spectroscopic coverage \citep{aadr1,aadr3,aadr5}.
Together with still unpublished data, ALFALFA has fully catalogued the
following two sky regions relevant for GASS: 
(a) $7.5< \alpha_{2000}< 16.5$ hrs, $+4^{\circ}< \delta_{2000}< +16^{\circ}$,
and (b) $22< \alpha_{2000}< 3$ hrs, $+14^{\circ}< \delta_{2000}< +16^{\circ}$.
Notice that the only ALFALFA detections that we do not re-observe are
the ones classified as
reliable (\ie, ``code 1'' in their tables); we do target the ALFALFA
{\em priors}, \ie\ candidate sources with lower
signal-to-noise ($<6.5$) but optical counterparts with known and
matching redshift (``code 2''). These sources are usually confirmed by
GASS observations and detected with short integrations (13 of the 176
DR1 galaxies are classified as ALFALFA code 2. They were all detected
except one).
In the rest of this paper, ``ALFALFA detections'' will refer only to
the galaxies in the first category.
Figure~\ref{ps_skyd} shows the sky distribution of the 769 sources
meeting GASS selection criteria that have been detected by the ALFALFA
survey to date (blue points). These are divided into 658 in the
region (a) defined above and 81 in (b), corresponding to 18\% and 19\%
of the parent sample in the same areas. The S05 archive
contributes 195 galaxies to region (a) and 45 to (b), the large
majority of which are also ALFALFA detections (but the S05
\hi\ profiles have higher signal-to noise). After removing
duplicates, the numbers of available \hi\ detections in regions (a) and
(b) are 710 and 89, respectively.

\section{Arecibo Observations and Data Reduction}\label{s_obs}

GASS observations started in March 2008 and are on-going. 
In the first 1.5 years of the survey until end of August 2009, Arecibo
allocated 239 hours to this project, with approximately 40 hours
lost to technical or RFI problems. 
The observations were scheduled in 85 blocks of 1$-$6.25 hours,
with about 50\% of the total allocation time in blocks of 3 hours or
less. The full survey will require a total of \about 900 hours to be
completed (not including the increased overhead caused by the
scheduling in short blocks).
Most of the observing (\about 160 hours) has been carried out
remotely from the MPA in Garching. Since May 2009, remote
observations are also done from Columbia, JHU, and NYU.

The \hi\ observations are done in standard {\em position-switching}
mode: each observation consists of an {\it on/off} source pair, each
typically integrated for 5 minutes, followed by the
firing of a calibration noise diode. We use the L-band wide receiver,
which operates in the frequency range 1120$-$1730 MHz, with a
1280$-$1470 MHz filter to limit the impact of RFI on our observations.
The interim correlator is used as a backend.
The spectra are recorded every second with 9-level sampling. 
Two correlator boards, each configured for 12.5 MHz
bandwidth, one polarization, and 2048 channels per spectrum
(yielding a velocity resolution of 1.4 \kms\ at 1370 MHz before
smoothing) are centered at or near the frequency
corresponding to the SDSS redshift of the target. The other two boards
are configured for 25 MHz bandwidth, two polarizations, 1024
channels per spectrum, and are centered at 1365 and 1385 MHz,
respectively. This setup allows us to monitor the full frequency
interval of the GASS targets (1353 to 1386 MHz, corresponding to
$z=0.050$ and $z=0.025$, respectively) for RFI and other problems.
All the observations are done during night-time to minimize the impact
of RFI and solar standing waves on our data.
The Doppler correction for the motion of the Earth is applied during
off-line processing.   

A radar blanker is used for \about 2/3 of our targets (those below
1375 MHz) to avoid RFI caused by harmonics of the FAA
airport radar in San Juan, which transmits at 1330 and 1350 MHz. This
device is synchronized with the pulsed signal of the radar --- the
data acquisition with the interim correlator is effectively
interrupted for a time $\delta t$ during and after each radar pulse
($\delta t$ can be chosen by the observer to lie between 100 and 750
$\mu$s, and is typically set to 400 $\mu$s to block, in addition to
the radar pulse, its delayed reflections from aircraft or other
obstacles). \\

The target lists for the observing runs are prepared in advance. The
selection is made from the compilation of 
galaxies mentioned in \S~\ref{s_sample}, only a small subset of which
are visible during each given observing window. The GASS targets are
randomly chosen from that subset, after excluding galaxies with prior
\hi\ detections from ALFALFA or the S05 archive, or with strong continuum 
sources within the beam that would cause ripples in the baselines.

We usually acquire up to three 5-minute {\it on-off}
pairs per object per observing session, and accumulate observations
from multiple sessions. One such pair requires approximately 13.5 or
11.3 minutes with or without the radar blanker, respectively, and
exclusive of slew-to-source time.
For galaxies with $T_{\rm max} \le 4$ minutes, or to use end-of-run
blocks of order of 10 minutes, we acquire 4-minute pairs instead, but
not shorter\footnote{
We considered using 2-minute integrations for galaxies
classified as ``code 2'' detections in ALFALFA (see
\S~\ref{s_overlap}), which has an effective integration time of 48
seconds. We targeted two of these objects during the very first run of
the survey -- one was detected (but the baseline was less than
optimal) and one was not.
}.
This is a good compromise between
obtaining good quality profiles with small time investment for
\hi-rich objects and minimizing overheads (which increase when an
observation is broken into smaller time segments). 
On-source integration times for the sample presented in this work
ranged between 4 and 90 minutes, with an average of 13 minutes for the
detections and 23 minutes for the non-detections (total times are
\about 2.5 times longer).

GASS spectra are quickly combined and
processed during the observations in order to assess data quality, and
to allow us to stop the integration when the object is detected, or when the maximum
time to reach its limiting gas fraction has been reached. A more
careful data reduction, which includes RFI excision, is performed
off-line at a later stage.\\

The data reduction is performed in the IDL environment using our own
routines, which are based on the standard Arecibo data processing
library developed by Phil Perillat. More specifically, we adapted
the software written to process and measure the observations described
in \citet{highz}, which, apart for targeting higher redshift objects,
adopted identical observing mode and setup.
In summary, the data reduction of each polarization and
{\it on-off} pair includes  Hanning smoothing, 
bandpass subtraction, RFI excision, 
and flux calibration. A total spectrum is obtained for each of the
two orthogonal linear polarizations by combining
good quality records (those without serious RFI or standing waves).
Each pair is weighted by a factor $1/rms^2$, where $rms$ is the root mean square
noise measured in the signal-free portion of the spectrum. 
The two polarizations are separately inspected (they usually agree
well. If present, polarization mismatches are noted in 
the Appendix), and averaged to produce the final spectrum.

After boxcar smoothing and baseline subtraction, the \hi-line
profiles are ready for the measurement of redshift, rotational
velocity and integrated  \hi\ line flux. 
Recessional and rotational velocities are measured at the 50\%
peak level from linear fits to the edges of the \hi\ profile.
Our measurement technique is explained in more detail, \eg, in
\citet[][\S 2.2]{widths}.

\section{SDSS and GALEX Data}\label{s_sdss}

This section summarizes the quantities derived from optical and UV
data used in this paper. All the optical parameters listed below were
obtained from Structured Query Language (SQL) queries to the SDSS
DR7 database server\footnote{
http://cas.sdss.org/dr7/en/tools/search/sql.asp
},
unless otherwise noted.

The GALEX UV photometry for our sample was completely reprocessed by
us, as explained in \citet{jing09}.
Briefly, we registered GALEX NUV/FUV and SDSS \rband\ images, and
convolved the latter to the (lower resolution) UV Point Spread
Function (PSF) using Image Reduction and Analysis Facility (IRAF)
tasks. The UV PSFs are measured from stacked stellar images, obtained by
co-adding the stars within 1200 pixels from the center of the frame. 
After masking out nearby sources detected in either UV or convolved
SDSS \rband\ images, we used SExtractor \citep{sextractor} to calculate magnitudes within
Kron elliptical apertures, defined on the convolved SDSS images.

The \nuvr\ colours thus derived are corrected for Galactic extinction
following \cite{wyder07}, who adopted $A(\lambda)/E(B-V)= 2.751$ for
the SDSS \rband\ and $A(\lambda)/E(B-V)=8.2$ for GALEX NUV. From these
assumptions, the correction to be applied to \nuvr\ colours is 
$A_{NUV}-A_r = 1.9807 A_r$, where the extinction $A_r$ is obtained
from the SDSS data base (listed in Table~\ref{t_sdss} below as ``$ext_r$'').

Internal dust attenuation corrections are very uncertain
for galaxies outside the blue sequence, especially in absence 
of far infrared data \citep[\eg,][]{johnson07,wyder07,luca08}.
Moreover, obtaining reliable SFRs from NUV photometry
is not trivial in the poorly calibrated, low specific SFR regime
\citep[\eg,][]{david07,salim07}. A detailed discussion of these
issues is beyond the scope of this work, and is presented in Paper II.
Hence, we do not correct \nuvr\ colours for dust attenuation, nor we derive
dust-corrected SFRs in this paper.\\

Table~\ref{t_sdss} lists the relevant SDSS and UV quantities for the GASS
objects published in this work, ordered by increasing right ascension:\\
Cols. (1) and (2): GASS and SDSS identifiers. \\
Col. (3): SDSS redshift, $z_{\rm SDSS}$. The typical uncertainty of
SDSS redshifts for this sample is 0.0002.\\
Col. (4): base-10 logarithm of the stellar mass, \Mst, in solar
units. Stellar masses are derived from SDSS photometry using the
methodology described in \citealt{salim07} (a \citealt{chabrier03}
initial mass function is assumed).
Over our required stellar mass range, these values are
believed to be accurate to better than 30\%, significantly smaller
than the uncertainty on other derived physical parameters such as star
formation rates.
This accuracy in \Mst\ is more than sufficient for this study.\\
Col. (5): radius containing 50\% of the Petrosian flux in \zband, \Rinz,
in arcsec.\\
Cols. (6) and (7): radii containing 50\% and 90\% of the Petrosian
flux in \rband, $R_{50}$ and  $R_{90}$ respectively, in arcsec (for
brevity, we omit the subscript ``$r$'' from these quantities
throughout the paper).\\
Col. (8): base-10 logarithm of the stellar mass surface density, \must, in
\Msun~kpc$^{-2}$. This quantity is defined as 
$\mu_\star = M_\star/(2 \pi R_{50,z}^2)$, with \Rinz\ in kpc units.\\
Col. (9): Galactic extinction in \rband, ext$_r$, in magnitudes, from SDSS.\\
Col. (10): \rband\ model magnitude from SDSS, $r$, corrected for Galactic extinction.\\
Col. (11): \nuvr\ observed colour from our reprocessed photometry,
corrected for Galactic extinction.\\
Col. (12): exposure time of GALEX NUV image, T$_{NUV}$, in seconds.\\
Col. (13): maximum on-source integration time, \tmax, required to
reach the limiting \hi\ mass fraction, in minutes (see \S~\ref{s_design}).
Given the \hi\ mass limit of the galaxy (set by its gas fraction limit
and stellar mass), we computed the required integration time to reach
this limit at the galaxy's redshift, assuming a 5$\sigma$ signal with
300 \kms\ velocity width and the instrumental parameters typical of
our observations (\ie, gain \about 10 K Jy\minusone\ and system
temperature \about 28 K at 1370 MHz).

\section{\hi\ Source Catalogs}\label{s_hi}

In this section we present the main \hi\ parameters of the 99 galaxies 
detected by GASS to date, and provide upper limits for the 77 objects that
were not detected.

Table~\ref{t_det} lists the derived \hi\ quantities for the detected
galaxies (ordered by increasing right ascension), namely:\\
Cols. (1) and (2): GASS and SDSS identifiers. \\
Col. (3): SDSS redshift, $z_{\rm SDSS}$, repeated here from
Table~\ref{t_sdss} to facilitate the comparison with the
\hi\ measurement (col.~6). \\
Col. (4): on-source integration time of the Arecibo
observation, $T_{\rm on}$, in minutes. This number refers to
{\it on scans} that were actually combined, and does not account for
possible losses due to RFI excision (usually negligible). \\
Col. (5): velocity resolution of the final, smoothed spectrum in \kms. \\
Col. (6): redshift, $z$, measured from the \hi\ spectrum.
The error on the corresponding heliocentric velocity, $cz$, 
is half the error on the width, tabulated in the following column.\\
Col. (7): observed velocity width of the source line profile
in \kms, \whi, measured at the 50\% level of each peak. 
The error on the width is the sum in quadrature of the 
statistical and systematic uncertainties in \kms. Statistical errors
depend primarily on the signal-to-noise of the \hi\ spectrum, and are
obtained from the rms noise of the linear fits to the edges of the
\hi\ profile. Systematic errors depend on the subjective choice of the
\hi\ signal boundaries, and are estimated as explained in
\citet{aadr1}. These are negligible for most of the galaxies in this
sample (only 17 objects have systematic errors greater than zero).\\
Col. (8): velocity width corrected for instrumental broadening
and cosmological redshift only, \whi$^c$, in \kms. No inclination
or turbulent motion corrections are applied.\\
Col. (9): observed, integrated \hi-line flux density in Jy \kms,
$F \equiv \int S~dv$, measured on the smoothed and baseline-subtracted
spectrum. The reported uncertainty is the sum in quadrature of the 
statistical and systematic errors (see col. 7).
The statistical errors are calculated according to equation 2 of S05. \\
Col. (10): rms noise of the observation in mJy, measured on the
signal- and RFI-free portion of the smoothed spectrum.\\
Col. (11): signal-to-noise ratio of the \hi\ spectrum, S/N,
estimated following \citet{saintonge07} and adapted to the velocity
resolution of the spectrum. 
This is the definition of S/N adopted by ALFALFA, which accounts for the
fact that for the same peak flux a broader spectrum has more signal.\\
Col. (12): base-10 logarithm of the \hi\ mass, \Mhi, in solar
units, computed via: 
\begin{equation}
    \frac{M_{\rm HI}}{\rm M_{\odot}} = \frac{2.356\times 10^5}{1+z}
    \left[ \frac{d_{\rm L}(z)}{\rm Mpc}\right]^2
    \left(\frac{\int S~dv}{\rm Jy~km~s^{-1}} \right)
\label{eq_MHI}
\end{equation}
\noindent
where $d_{\rm L}(z)$ is the luminosity distance to the galaxy at
redshift $z$ as measured from the \hi\ spectrum. \\
Col. (13): base-10 logarithm of the \hi\ mass fraction, \Mhi/\Mst.\\
Col. (14): quality flag, Q (1=good, 2=marginal, 5=confused). 
An asterisk indicates the presence of a note for the source in the Appendix.
Code 1 refers to reliable detections, with a S/N ratio of order of
6.5 or higher (this is the same threshold adopted by ALFALFA). 
Marginal detections have lower S/N, thus more uncertain
\hi\ parameters, but are still secure detections, with \hi\ redshift
consistent with the SDSS one.
The S/N limit is not strict, but depends also on \hi\ profile and baseline
quality. As a result, galaxies with S/N slightly above the threshold
but with uncertain profile or bad baseline may be flagged with a
code 2, and objects with S/N $\lesssim 6.5$ and \hi\ profile with
well-defined edges may be classified as code 1. 
We assigned the quality flag 5 to four ``confused'' galaxies, where
most of the \hi\ emission is believed to come from another source
within the Arecibo beam. For some of the galaxies, the presence of
small companions within the beam might contaminate (but is unlikely to
dominate) the \hi\ signal -- this is just noted in the Appendix.
Finally, we assigned code 3 to GASS 9463, which is both marginal and
confused.\\

Table~\ref{t_ndet} gives the derived \hi\ upper limits for the non-detections. 
Columns (1-4) and (5) are the same as columns (1-4) and (10) in Table~\ref{t_det},
respectively. Column (6) lists the upper limit on the \hi\ mass in
solar units, Log \Mhi$_{,lim}$, computed assuming a 5 $\sigma$ signal with 300 \kms\ 
velocity width, if the spectrum was smoothed to 150 \kms. Column (7)
gives the corresponding upper limit on the gas fraction, Log \Mhi$_{,lim}$/\Mst.   
An asterisk in Column (8) indicates the presence of a note for the
galaxy in the Appendix.\\

\begin{figure*}
\includegraphics[width=16.5cm]{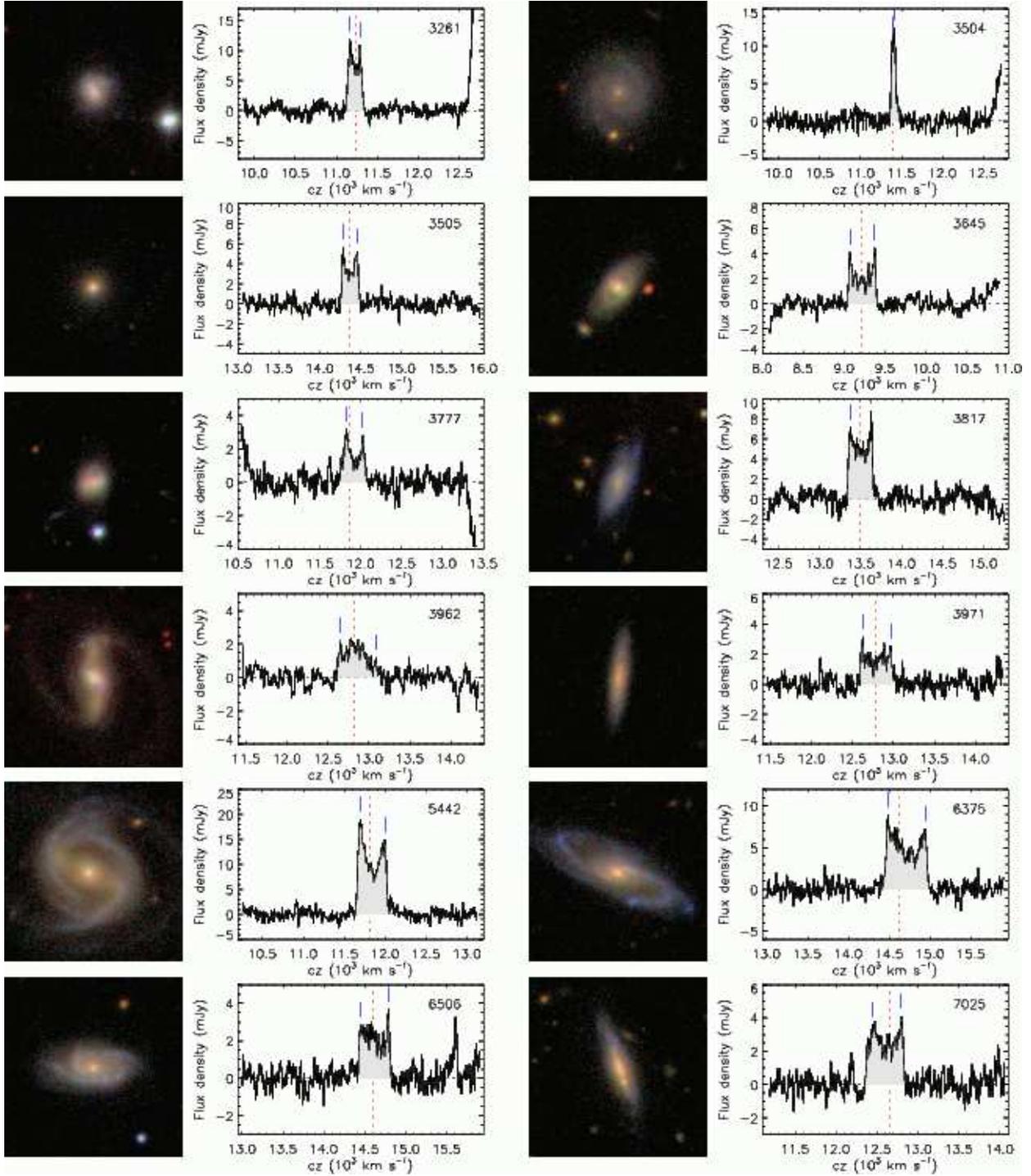}
\caption{SDSS postage stamp images (1 arcmin square) and
  \hi-line profiles of the survey detections, ordered by increasing GASS
  number (indicated in each spectrum). The \hi\ spectra are
  calibrated, smoothed and baseline-subtracted. A dotted line and two
  dashes indicate the heliocentric velocity corresponding to the SDSS
  redshift and the two peaks used for width measurement, respectively.
[{\em See the electronic edition of the Journal for the complete figure.}]}
\label{det}
\end{figure*}

\begin{figure*}
\includegraphics[width=16.5cm]{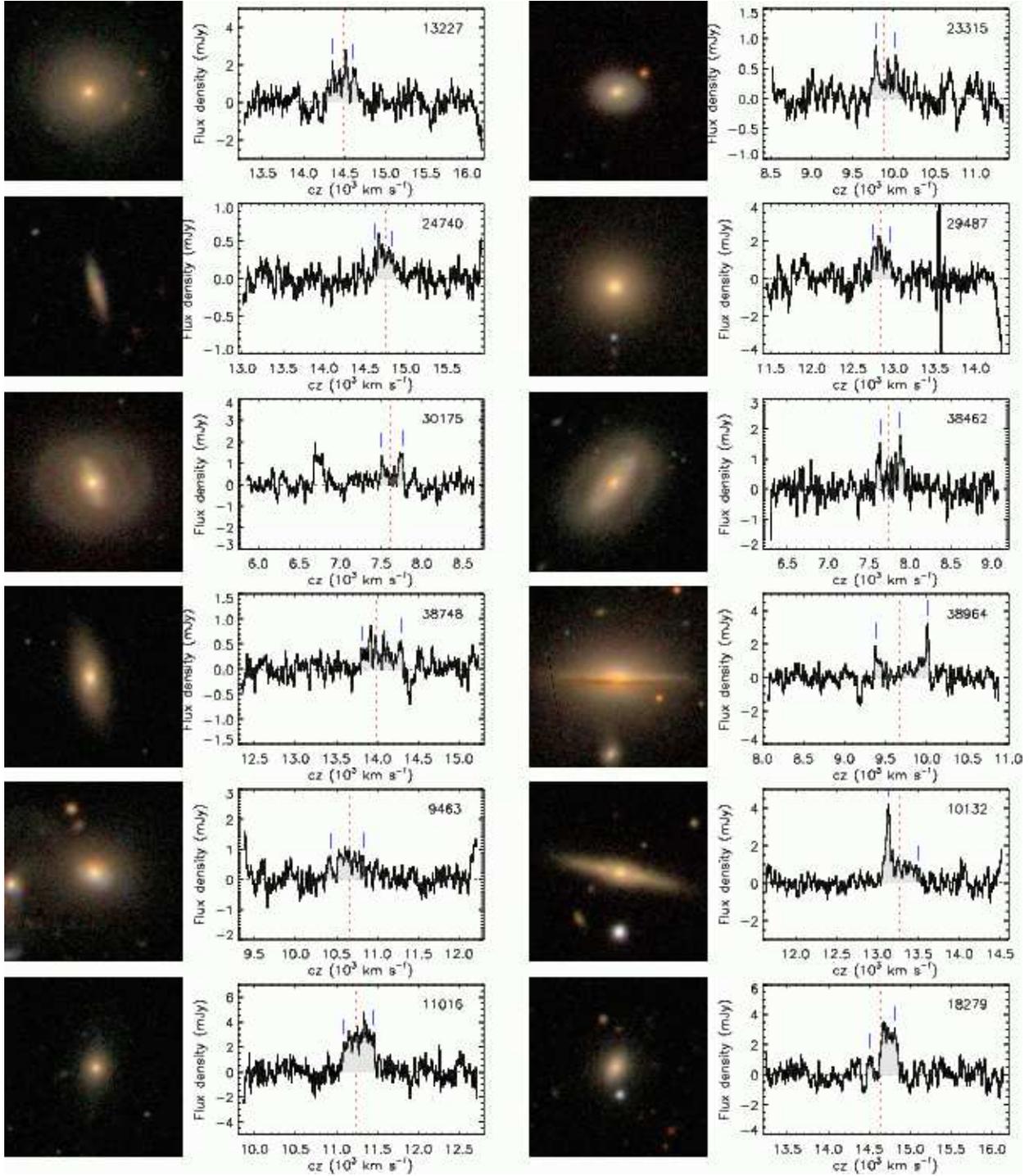}
\caption{Same as Fig.~\ref{det} for marginal (first four rows) or
confused (last two rows; GASS 9463 is both marginal and confused) detections.
[{\em See the electronic edition of the Journal for the complete figure.}]}
\label{marg}
\end{figure*}

\begin{figure*}
\includegraphics[width=16.2cm]{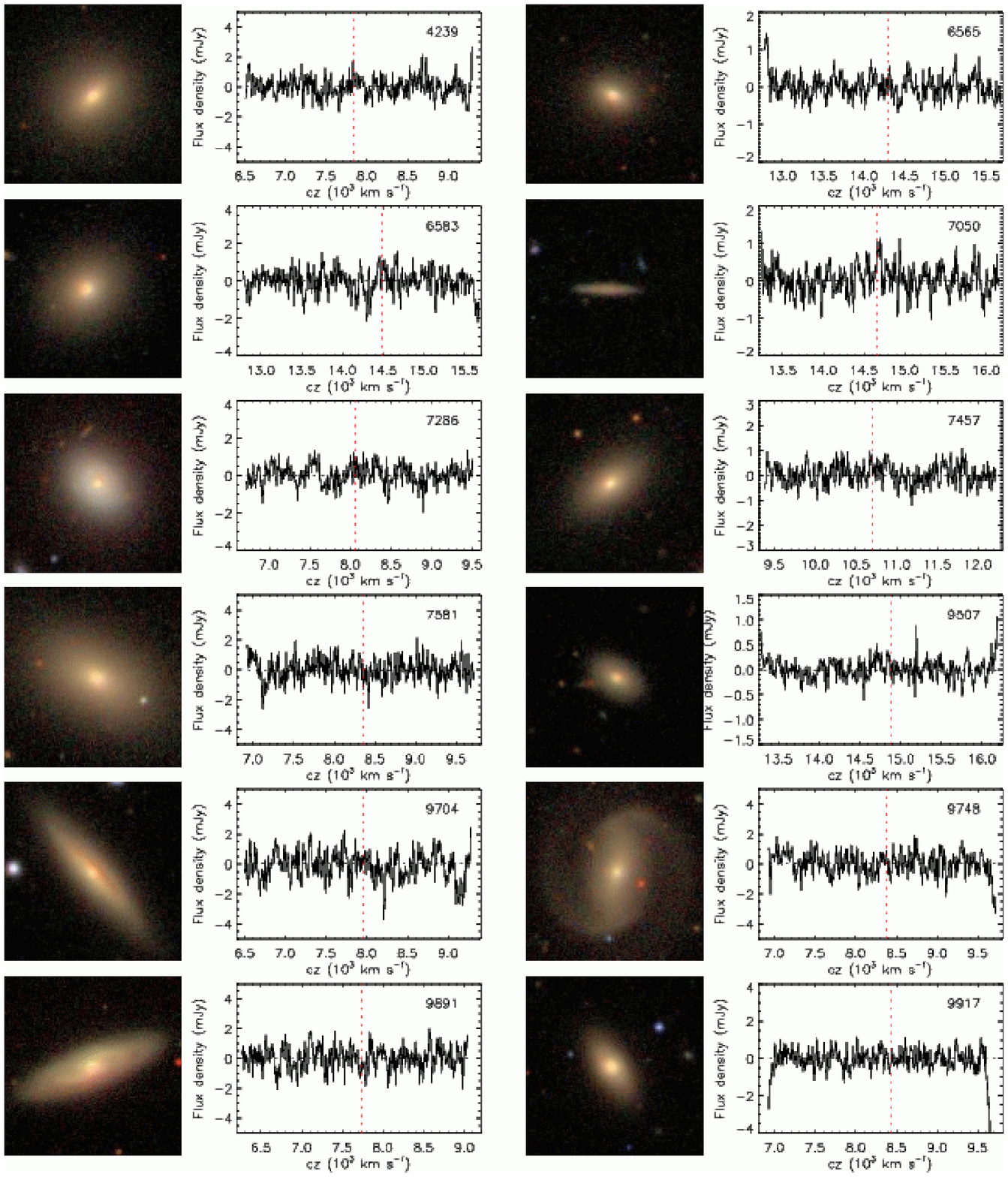}
\caption{Same as Fig.~\ref{det} for non-detections.
[{\em See the electronic edition of the Journal for the complete figure.}]}
\label{ndet}
\end{figure*}

Figure~\ref{det} shows SDSS images and \hi\ spectra for the galaxies
with quality flag 1 in Table~\ref{t_det}; marginal detections and galaxies for which 
confusion is certain are shown separately in Figure~\ref{marg}, and non-detections
are presented in Figure~\ref{ndet}. The objects in these figures are ordered by 
increasing GASS number (indicated on the top right corner of each spectrum).
The SDSS images show a 1 arcmin square field, \ie, only the central
part of the region sampled by the Arecibo beam (the half
power full width of the beam is \about 3.5\arcmin\ at the
frequencies of our observations). Therefore, companions that might be
detected in our spectra typically are not visible in the
postage stamps -- examples include GASS 40007, a spectacular pair of blue
spirals with 1.3\arcmin\ separation and 60 \kms\ velocity difference
(marked as confused in Table~\ref{t_det}, even if the \hi\ spectrum
does not appear clearly distorted), and a few non-detections (\eg,
GASS 29090, 40686, and 42156).
The \hi\ spectra are always displayed over a 3000 \kms\ velocity
interval, which includes the full 12.5 MHz bandwidth adopted for our
observations. The \hi-line profiles are calibrated, smoothed 
(to a velocity resolution between 5 and 21 \kms\ for
the detections, as listed in Table~\ref{t_det}, or to
\about 15 \kms\ for the non-detections), and
baseline-subtracted. A red, dotted line indicates the heliocentric
velocity corresponding to the optical redshift from SDSS. There is
a very good agreement between SDSS and \hi\ redshifts, with only small
offsets that are usually within the typical SDSS measurement
uncertainty (0.0002).
In Figures~\ref{det} and \ref{marg}, the shaded area and two vertical
dashes show the part of the profile that was integrated to
measure the \hi\ flux and the peaks used for width measurement, respectively.

The GASS \hi\ spectral data products will be incorporated into the
Cornell \hi\ digital archive\footnote{
{\em http://arecibo.tc.cornell.edu/hiarchive}
}, a registered Virtual Observatory node that already contains the ALFALFA data
releases and the S05 \hi\ archive of targeted observations.

\begin{figure*}
\includegraphics[width=14cm]{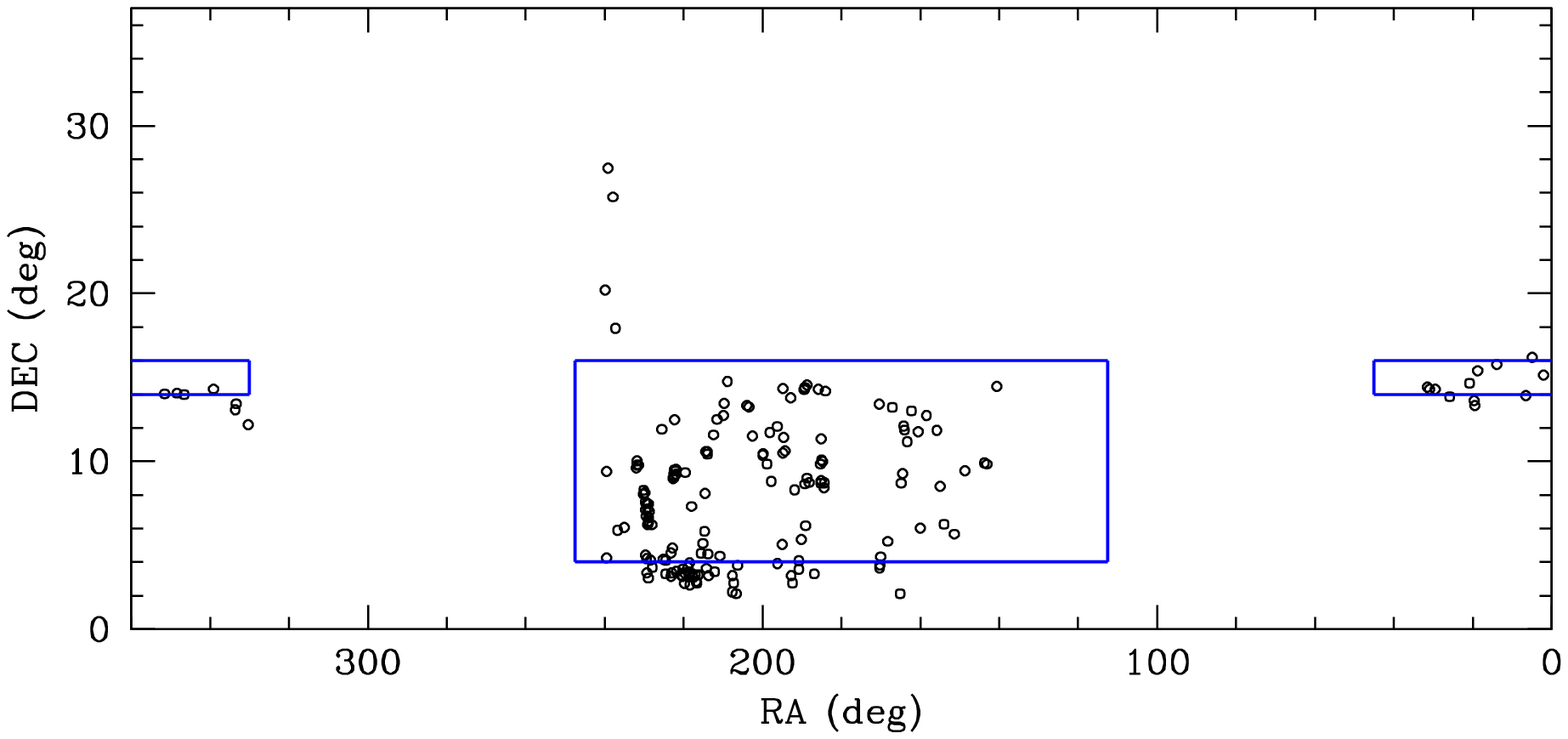}
\caption{Sky distribution of the 176 galaxies in the GASS Data Release 1. 
The rectangles indicate regions already catalogued by ALFALFA to date 
(ALFALFA coverage of areas without SDSS spectroscopy is not shown).}
\label{skyd}
\end{figure*}

\section{Results}\label{s_res}

\subsection{GASS DR1 Sample}\label{s_dr1}

We describe here the properties of the galaxies included in this first 
GASS data release, whose Arecibo \hi-line profiles and derived
parameters were presented in the previous section.

The sky distribution of the galaxies is shown in Figure~\ref{skyd}. 
As can be seen, most of the observing time thus far has been allocated
in the $110^{\circ}< \alpha_{2000}< 250^{\circ}$ region (referred to
as the ``Spring sky'', because  visible from Arecibo during the
night-time in that season). The targets are also concentrated in the
part of sky already covered and catalogued by ALFALFA (rectangles). 
We did observe several galaxies located outside the current ALFALFA
catalogued footprint, partly because of time allocation constraints and partly
because, as already mentioned, we gave some priority to objects with
stellar mass larger than $10^{10.5}$ \Msun.

The distributions of measured \hi\ properties for GASS detections are
presented in Figure~\ref{dr1hi} (solid histograms). In the top left
panel, we show the redshift histogram for the full DR1 sample using
SDSS measurements (dotted). The distribution of
velocity widths (not deprojected to edge-on view) peaks near 300 \kms.
As mentioned in the previous section, this is the value we adopt to
calculate \hi\ mass limits for the non-detections. The measured
\hi\ masses vary between $4.6 \times 10^{8}$ and $3.2 \times 10^{10}$ \Msun. 
Approximately half of the galaxies detected by GASS have \hi\ mass
fractions smaller than 10\%.

\begin{figure*}
\includegraphics[width=12cm]{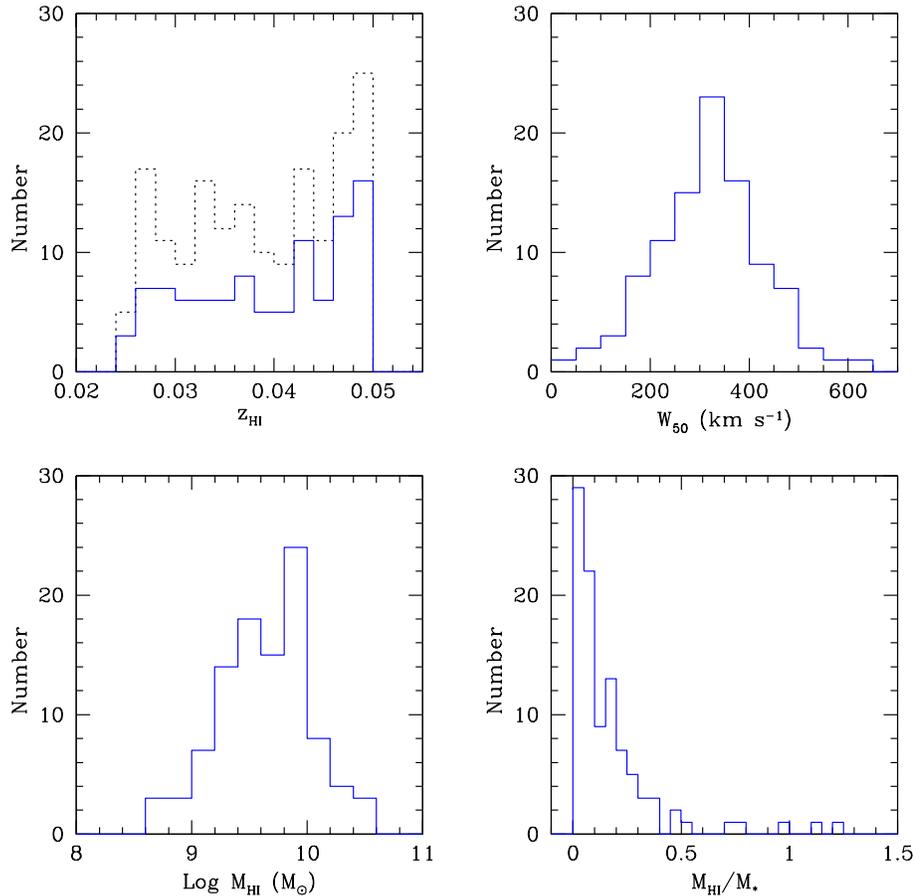}
\caption{Distributions of redshifts, velocity widths,
\hi\ masses and gas mass fractions for the 99 galaxies with
\hi\ detections. The dotted histogram in the top left panel shows the
distribution of SDSS redshifts for the full DR1 sample (\ie, including
the non-detections).}
\label{dr1hi}
\end{figure*}

\begin{figure*}
\includegraphics[width=12cm]{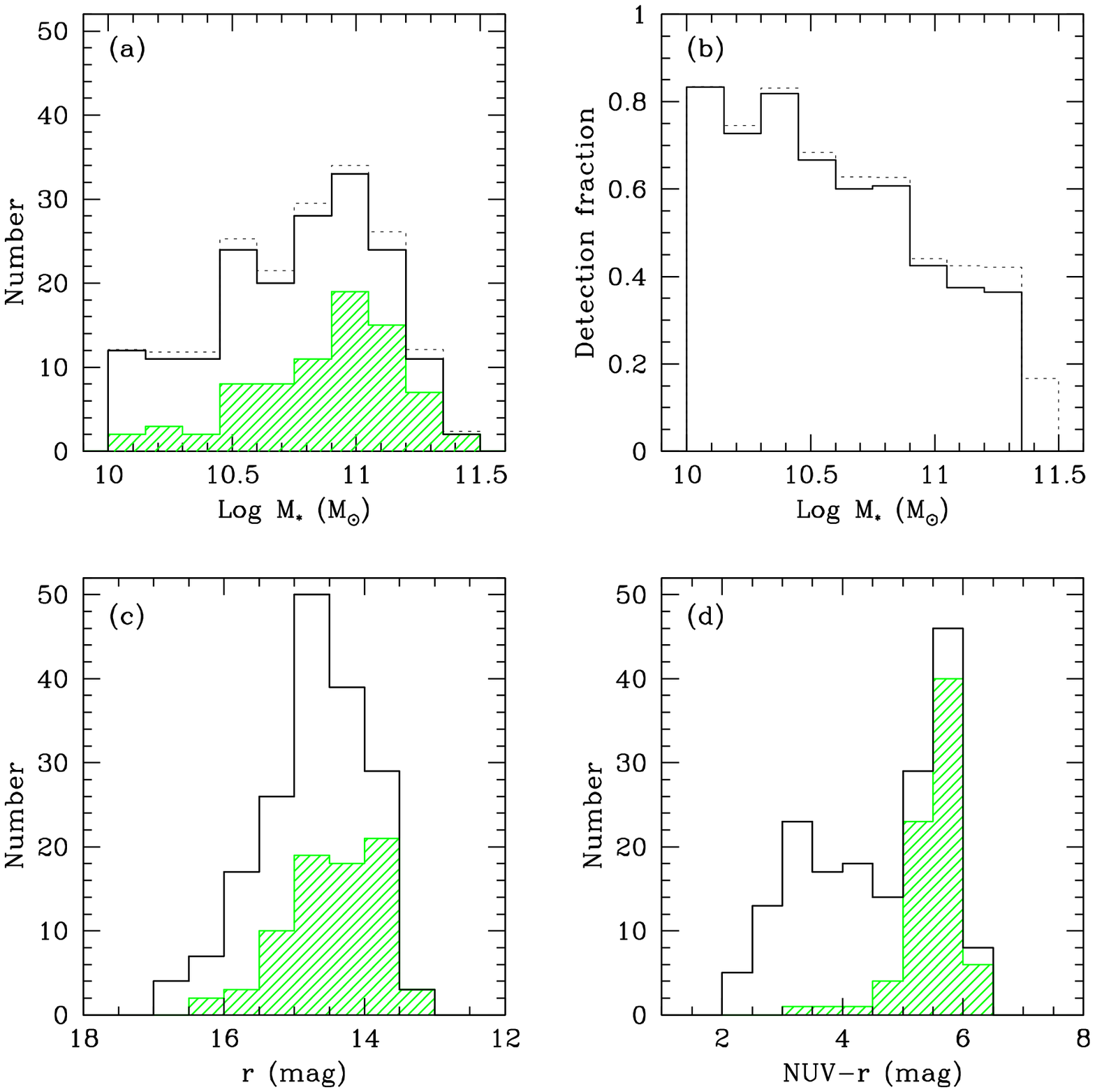}
\caption{Stellar mass (a), SDSS \rband\ (c) and observed \nuvr\ colour (d)
distributions for the GASS DR1 sample (solid). Hatched
histograms indicate the corresponding distributions for the non-detections.
The detection fraction (\ie, the ratio of detections to total) is
shown as a function of stellar mass in (b). The dotted histograms in
the top panels are the {\em average} distributions for the
representative sample discussed in the text (\S~\ref{s_repr}).
}
\label{dr1hist}
\end{figure*}

An overview of some of the optical- and UV-derived parameters for this
data set is found in Figure~\ref{dr1hist}, where solid and
hatched histograms represent full sample and non-detections, respectively
(dotted histograms will be discussed in the next section).
Galaxies that were not detected in \hi\ are spread throughout the
entire GASS stellar mass interval (panel a), but are preferentially
found at higher \Mst\ values. The detection rate of the survey as a
function of \Mst\ is shown in panel (b). The fraction of
detected galaxies decreases with stellar mass, but does not
drop below 35\% even at stellar masses larger than $10^{11}$ \Msun\ 
(except for the very last bin, where we targeted only two objects).
Thus, GASS is sensitive enough to detect \hi\ in a
significant fraction of massive systems.
The bottom panels of Figure~\ref{dr1hist} show SDSS \rband\ magnitude
and observed (\ie, not corrected for dust attenuation)    
\nuvr\ colour distributions for this sample. The well-known
separation between blue cloud and red sequence galaxies, best
appreciated when UV-to-optical colours are used
\citep[\eg,][]{wyder07}, is clearly seen in
panel (d). Not surprisingly, nearly all the non-detections are found
in the red sequence, whereas bluer, star-forming objects are almost
always detected. Particularly interesting are the galaxies detected in
red sequence. Of the 17 detections with \nuvr $>5$, half are highly 
inclined disks, thus their colours are likely reddened by dust, but
most of the others have a featureless, spheroidal appearance in the SDSS
images. The gas fractions of these detections are all below 6\%, with
two very notable exceptions, GASS 9863 and 3505 (\hi\ mass fractions
of 27\% and 50\%, respectively), both early-type galaxies. GASS 3505
in particular is an extraordinary system that will be mentioned again
in this work --- its SDSS image and Arecibo detection can be seen in
the left column of the second row of Figure~\ref{det}.

\begin{figure*}
\includegraphics[width=17cm]{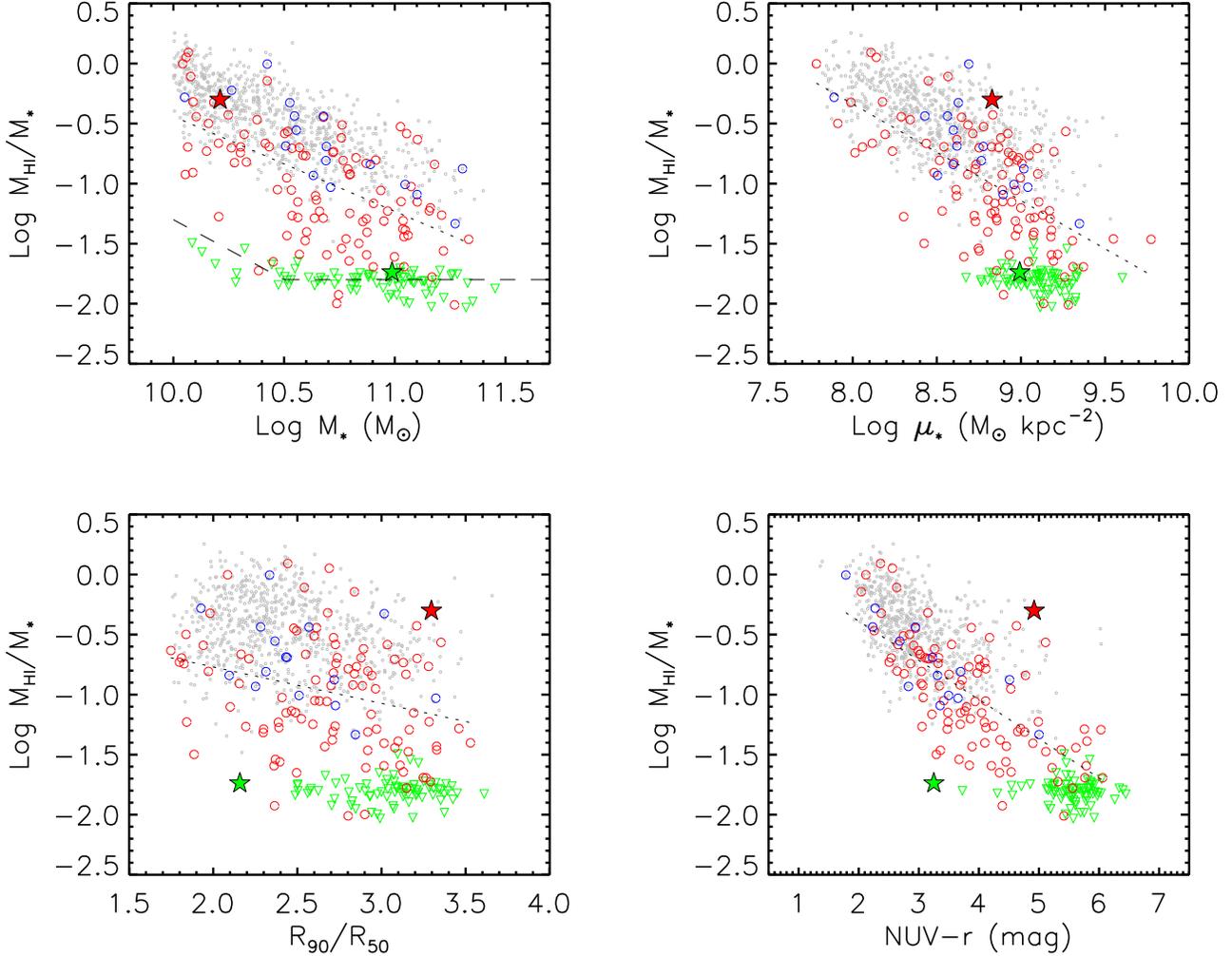}
\caption{The \hi\ mass fraction of the GASS sample is plotted here as
a function of stellar mass, stellar mass surface density,
concentration index, and observed \nuvr\ colour. Red circles and green triangles
represent detections and non-detections, respectively. 
Blue symbols indicate randomly selected, \hi-rich ALFALFA and S05
galaxies added to the DR1 sample in the correct proportion, to
compensate for the fact that we do not re-observe objects with good
detections in either of those archives. The sample including 
red, green and blue symbols is one example of the catalogs
described in \S~\ref{s_repr}, and contains 193 galaxies.
For comparison, we also show the full set of ALFALFA galaxies meeting the GASS
selection criteria that have been catalogued to date (gray). The stars
in all plots highlight the locations of two interesting objects, GASS
3505 (red) and GASS 7050 (green), discussed in the text.
The dashed line on the top-left panel indicates the \hi\ detection limit
of the GASS survey. Dotted lines in each panel are linear fits to
the red and blue circles only.}
\label{dr1gf}
\end{figure*}

In Figure~\ref{dr1gf}, we examine how the gas mass fraction of this sample 
varies as a function of several other galaxy properties. From top to
bottom and from left to right, the \hi-to-stellar mass ratio is
plotted as a function of stellar mass, stellar mass density,
concentration index, and observed \nuvr\ colour.
In all panels, GASS detections are represented by red circles,
and upper limits for non-detections are plotted as green, upside-down triangles.
For comparison, the \hi-rich ALFALFA detections\footnote{
The \hi\ masses of galaxies in the ALFALFA and S05 archives have been
recomputed from the tabulated fluxes using Eq.~\ref{eq_MHI} and
adopting the cosmological parameters listed in \S~\ref{s_intro}.
}
in the GASS parent sample 
are shown as small gray circles. Blue circles are
randomly-selected, gas-rich galaxies from either ALFALFA or the S05
archive of pointed observations, added to the DR1 sample in the right
proportion to quantify average trends in the data (see \S~\ref{s_repr}).
The stars identify the locations of GASS 3505 (red) and GASS 7050
(green), which are somewhat extreme examples of two opposite types of transition
galaxies. The former is an early-type galaxy with unusually high gas
content that might be reaccreting a disk (thus moving from the red
toward the blue sequence), and the latter is a gas-poor, disk galaxy 
(which is likely moving in the opposite direction, from the blue to
the red sequence. The SDSS image and \hi\ spectrum of GASS 7050 can be seen in
the right column of the second row of Figure~\ref{ndet}).
This figure illustrates how GASS is able to measure the 
\hi\ content of massive galaxies down to a gas fraction limit which
is an order of magnitude lower than that achieved by ALFALFA over the same redshift range. 
This allows us to quantify the distribution of atomic gas fractions 
over a much wider dynamic range. The figure also illustrates
the relative strengths of the correlations (or lack thereof) between
gas content and galaxy structural parameters, and also between gas
content and \nuvr, a quantity that characterizes the stellar
populations of the galaxies in our sample.  

We discuss each panel below. In order to quantify the scatter in each
correlation, we have performed a least-squares fit to all the galaxies with 
\hi\ detections (\ie, all the points plotted as red and blue circles in Fig.~\ref{dr1gf}).
The result of the fit is shown as a dotted line in each panel. 
We only report the rms variance in Log~\Mhi/\Mst\ 
about each relation, $\sigma$, in the text below.

{\bf Stellar mass.} --- The apparently clean correlation between gas mass
fraction and \Mst\ exhibited by the \hi-rich ALFALFA galaxies
in this stellar mass regime is mostly due to the sensitivity limits of
the blind \hi\ survey. At each stellar mass, GASS detects many objects
with significantly lower gas fractions. The gas fraction of the GASS
detections decreases as a function of stellar mass, but the
correlation has larger scatter ($\sigma = 0.386$ dex) compared to the one obtained for the  
ALFALFA galaxies. As already noted, the non-detections span the full
\Mst\ interval studied by GASS, but their fraction of the targeted
sample increases with stellar mass. 
The stellar mass-dependent detection limit of our survey is indicated
by a dashed line
\footnote{
Several data points scatter below the nominal
detection threshold, and some non-detections lie slightly above it.
There are two main reasons for this. First, the dashed line is
the expected limit, computed assuming an average value for the 
telescope gain and a 5 $\sigma$ signal of fixed velocity width 
(300 \kms, which is representative for the massive galaxies that we
target). For the same \hi\ flux, galaxies with narrower profiles
(\ie, smaller {\em observed} velocity widths, which might be
intrinsic and/or due to small inclination to the line-of-sight) are
easier to detect, thus they might yield gas fractions below the dashed
line. Non-detections can scatter around that line because 
upper limits are based on the actual rms noise measured from the
spectra, which might differ slightly from the expected value (the
telescope gain depends on azimuth, zenith angle, and frequency of the
observation; the actual rms depends on baseline quality). 
Second, we never integrate less than 4 minutes, even if the maximum
time \tmax\ computed to reach the gas fraction limit is smaller 
(see \S~\ref{s_obs}). This affects the higher stellar mass
galaxies, for which the limit can be reached in as little as one
minute. As for the detections with small \tmax\ values, we consider
taking an additional {\em on/off pair} when the small investment of time
guarantees a significantly improved \hi\ profile.}.

{\bf Stellar mass density.} --- Interestingly, the gas content of
massive galaxies seems to correlate better with \must\ than with
\Mst. Quantitatively, the rms variance in Log~\Mhi/\Mst\ at a fixed value
of \must\ is $\sigma = 0.364$ dex, \ie\ 6\% smaller than the variance at a fixed value of \Mst.
Another striking feature in this plot is the
distribution of the non-detections, which all have 
\must $>10^{8.6}$ \Msun~kpc$^{-2}$, without a single exception.
We will come back to this point later.

{\bf Concentration index.} --- The concentration index is defined
as $R_{90}/R_{50}$, where $R_{90}$ and $R_{50}$ are the radii enclosing 90\% and 50\%
of the \rband\ Petrosian flux, respectively (see Table~\ref{t_sdss}). As shown in Figure 1 of
\citet{weinmann09}, there is a tight and well-defined
correlation between the concentration index and the bulge-to-total ratio
derived from full 2-dimensional multi-component fits using the methods
presented in \citet{gadotti09}. It is thus very interesting that 
the dependence of 
\Mhi/\Mst\ ratio on concentration is considerably weaker 
($\sigma = 0.449$ dex) than its dependence on
stellar mass or stellar surface density. Because the \hi\ gas is expected to be
located in the disk and not the bulge, this lack of correlation might imply that     
the bulge has little influence on the formation or fuelling of the disk.

{\bf NUV$-$r colour.} --- Colour is well known to be a reasonably good
predictor of gas content {\em for blue-sequence, star-forming galaxies}.
We show here that the correlation between those two quantities 
{\em continues} with increased scatter beyond the blue sequence traced
by the \hi-rich ALFALFA galaxies, and down to the survey gas
fraction limit. This is in agreement with previous work by,
\eg, \citet{luca09}, based on a sample of galaxies located in regions 
in and around nearby clusters.
Among the four parameters considered here (\ie, \Mst, \must, $R_{90}/R_{50}$
and \nuvr), \nuvr\ colour is the one most tightly correlated with gas
fraction ($\sigma = 0.327$ dex).
As seen in Figure~\ref{dr1hist}d, almost
all the non-detections are found in the red sequence.
We also notice that GASS 3505 and GASS 7050
stand out as clear outliers in this plot, with a gas content well
displaced from the average for their \nuvr\ colour.

In the remainder of the paper,  we will quantify the main correlations
explored in this section and discuss their implications.

\subsection{Building a Representative Sample}\label{s_repr}

Although this first data release amounts to only \about 20\% of the
full survey sample, our target selection has not been biased by any
galaxy property except stellar mass. As described in \S~\ref{s_sample}, in
this initial phase of the survey we have given some priority to
galaxies with stellar masses greater than $10^{10.5}$ \Msun, but
otherwise the selection has been random. We can thus use this sample
to quantify the correlations between \hi\ mass fraction and other
physical parameters discussed above, after accounting for the missing
\hi-rich objects that were not observed because they are in
ALFALFA or in the S05 archive.

In order to construct a sample of galaxies that is representative in
terms of \hi\ properties, we clearly need to add the correct
proportions of ALFALFA and S05 galaxies to the GASS data set.
We determined such proportions for each stellar mass bin as follows.
We used the ALFALFA catalog footprint available to date to estimate
the fractions of GASS parent sample galaxies detected by ALFALFA (\faa)
and in the S05 archive (\fhiar, not counting those already in \faa). 
One complication is that we observed targets outside that footprint,
some of which might turn out to be ALFALFA detections, and we should
not count these twice.
Thus, let us call \ngass\ and \ngassnorich\ the total number of GASS
objects in the given stellar mass bin and those below the
ALFALFA detection limit, respectively. The numbers of ALFALFA and S05
galaxies that should be added back into the sample, for each stellar
mass bin, are:
\[
N_{\rm AA} = \frac{N_{\rm G,<AA}}{1- f_{\rm AA}} \times f_{\rm AA} -(N_{\rm G}-N_{\rm G,<AA}),
\]
where ($N_{\rm G}-N_{\rm G,<AA}$) effectively count as ALFALFA detections
(they would be if that survey was already completed), and
\[
N_{\rm S05} =\frac{N_{\rm G,<AA}}{1- f_{\rm S05}} \times f_{\rm S05},
\]
where, as noted above, \fhiar\ does {\em not} include ALFALFA detections.
Repeating this process for each stellar mass bin yields a sample of
\about 200 galaxies. 

In order to better account for statistical fluctuations, 
we generated 100 such catalogs, each time selecting objects randomly
from the ALFALFA and S05 data sets in the correct proportions. More specifically,
each catalog is obtained by adding \naab\ and \nhiarb\ galaxies
to each \Mst\ bin of the GASS data set, where \naab\ and \nhiarb\ are  
random Poisson deviates of \naa\ and \nhiar, respectively. The
catalogs contain between 179 and 193 galaxies (187 on average), thus
the percentage of \hi-rich galaxies added to the GASS sample is less
than 10\%. This is illustrated by the blue symbols in
Figure~\ref{dr1gf} for one of these catalogs (red and green symbols
are GASS detections and non-detections, respectively, and blue circles
are randomly-selected \hi-rich galaxies that were added to each \Mst\ bin). 

When we compute the average \hi\ gas fraction as a function of different
galaxy properties, we do so based on the 100 catalogs discussed
here. The average data set obtained from these catalogs will be
referred to henceforth as our {\em representative sample}.   
To estimate  error bars, we also tried to take into account the
variance internal to the GASS sample itself using standard  
bootstrapping techniques. We generated 100 random catalogues in a
similar way to that described above, 
but we also draw random indices for GASS galaxies  
(allowing for repetitions). The catalogs in this second set
include between 179 and 200 galaxies each (189 on average).

\begin{figure*}
\includegraphics[width=17cm]{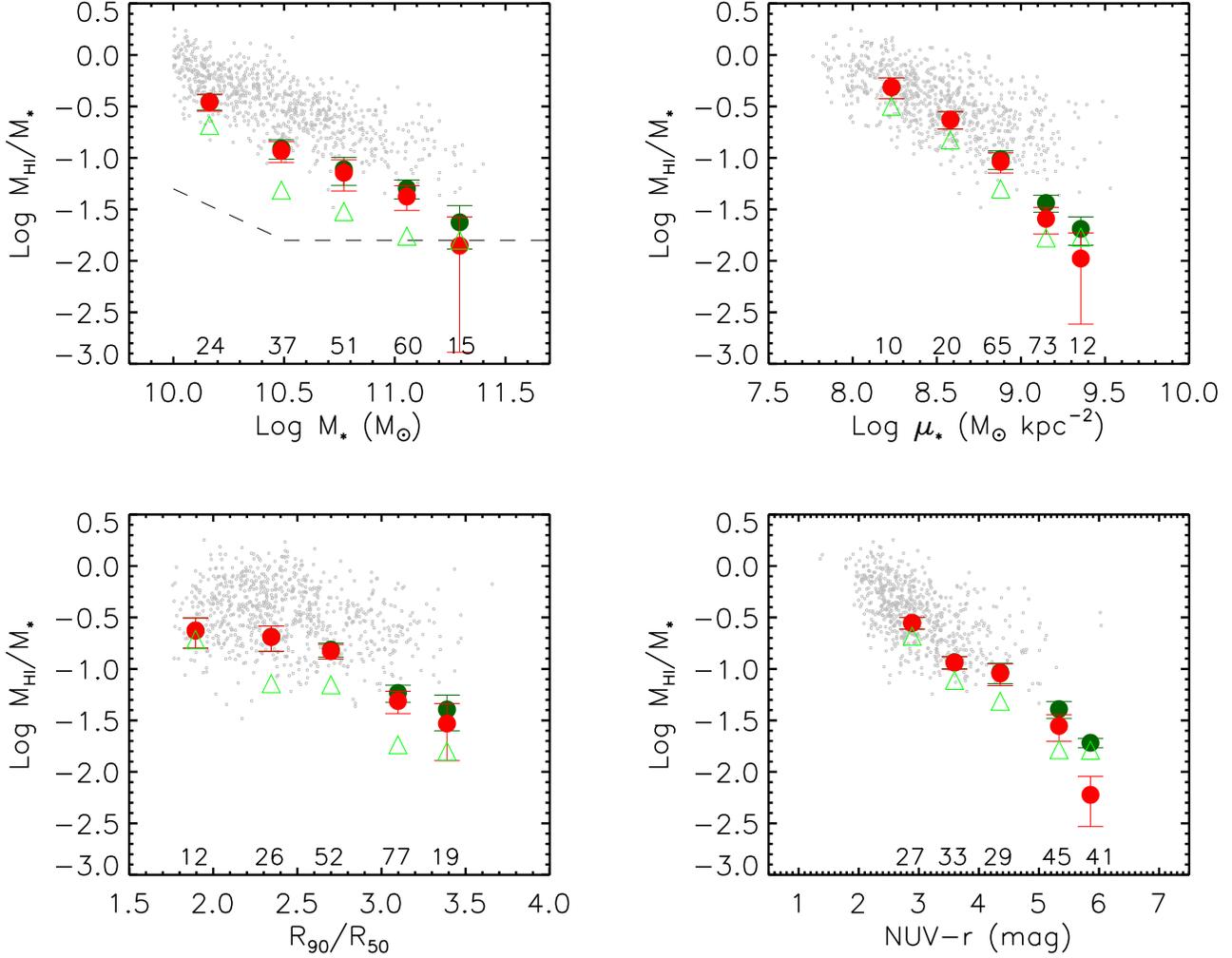}
\caption{Average trends of \hi\ mass fraction as a function of stellar
mass, stellar mass surface density, concentration index and observed
\nuvr\ colour based on the representative sample discussed in \S~\ref{s_repr}. 
In each panel, large circles indicate average gas fractions 
(weighted as explained in the text). These were computed including the
non-detections, whose \hi\ mass was set to either its upper limit
(dark green) or to zero (red). Green triangles are weighted medians.
The average number of galaxies in each bin, $N_{\rm i}$, is indicated
above the x axis; only bins with  $N_{\rm i} \ge 8$ are
shown. These results are listed in Table~\ref{t_avgs}. 
Galaxies in the GASS parent sample detected by ALFALFA are
plotted as small gray circles. The dashed line in the first panel
shows the \hi\ detection limit of the GASS survey.}
\label{scalings}
\end{figure*}

\subsection{Gas Fraction Scaling Relations}\label{s_gf}

In Figure~\ref{scalings}, we show how the average 
\hi\ mass fraction of massive galaxies varies as a function of stellar
mass, stellar mass surface density, concentration index and observed
\nuvr\ colour. This quantity is calculated for each of the 100
catalogs described in the previous section that do not include bootstrapping
of the GASS galaxies. We average the gas fractions
in a given bin, properly weighted in order to compensate for the
uneven stellar mass sampling of the DR1 data set, using the full GASS
parent sample as a reference.  We  bin both the parent sample
and the individual catalogs by stellar mass (with a 0.2 dex step), and
use the ratio between the two histograms as a weight. In other
words, a galaxy in the $i$-th stellar mass bin and in a given catalog
is assigned a weight $N_{{\rm PS},i}/N_{{\rm cat},i}$, where 
$N_{{\rm PS},i}$ and $N_{{\rm cat},i}$ are the numbers of objects in
the $i$-th Log~\Mst\ bin in the parent sample and in that catalog,
respectively. This is a valid procedure because the parent sample
is a {\em volume-limited} catalogue of galaxies with \Mst $> 10^{10}$ \Msun.
The final average gas fractions are then computed by averaging the weighted
mean values obtained for each catalog. The results are shown as large circles in 
Figure~\ref{scalings}. The difference between green and red circles illustrates
two different procedures for dealing with the non-detections: the 
\hi\ mass is set either to the upper limit (green) or to zero (red).
As can be seen, the answer is insensitive to the way we treat
the galaxies without \hi\ detections, except for the very most massive,
dense and red galaxies. 
Note that this is entirely due to the deep limit of the GASS survey.
The error bar on each circle indicates the 1$\sigma$ uncertainty on
our estimate of the average gas mass fraction, computed from the
dispersion of the weighted mean values of the bootstrapped catalogs.
The average number of galaxies that contributed to each bin is
indicated in the panels. Because of the small number statistics in
some of these bins, our bootstrapping analysis might still
underestimate the error bars.

\begin{figure*}
\includegraphics[width=14cm]{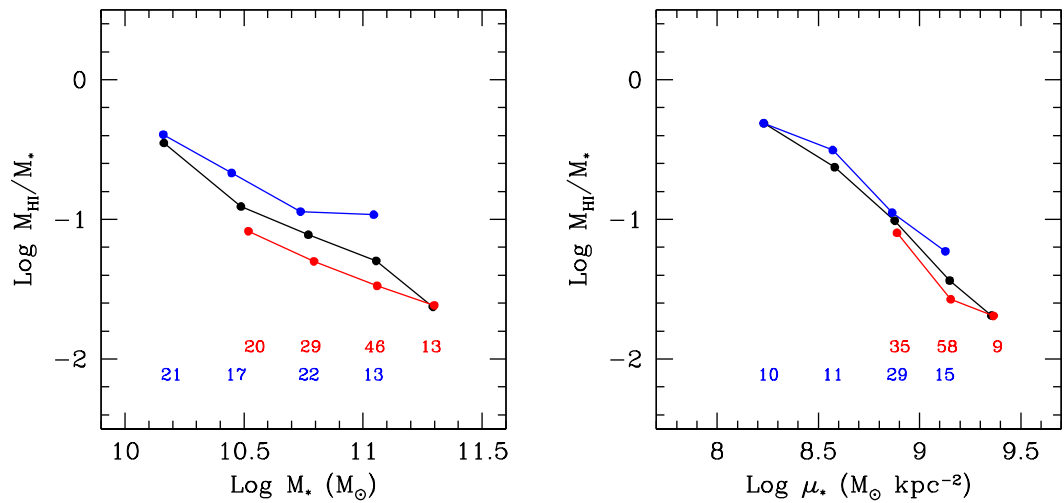}
\caption{Gas fraction as a function of stellar mass and stellar mass
density for our representative sample. Black symbols and lines are
weighted averages obtained by setting the non-detections to their
upper limits (same as dark green trends in Fig.~\ref{scalings}).
{\em Left:} The sample is divided into two bins of stellar mass
density (blue: Log \must $< 8.9$, red:  Log \must $\ge 8.9$). 
{\em Right:} The sample is divided into two bins of stellar mass
(blue: Log \Mst $< 10.7$, red:  Log \Mst $\ge 10.7$). Average numbers
of galaxies in each bin are indicated in the panels. Only averages
based on $N \ge 9$ galaxies are plotted. 
}
\label{split}
\end{figure*}

Weighted median\footnote{
Given n elements $x_1 ... x_n$ with positive weights $w_1 ... w_n$ such that
their sum is 1, the weighted median is defined as the element $x_k$ for which:
$\displaystyle \sum_{x_i < x_k} w_i < 1/2$ and 
$\displaystyle \sum_{x_i > x_k} w_i \leq 1/2$.
}
 values of the \Mhi/\Mst\ ratios are plotted in
Figure~\ref{scalings} as triangles. In the highest \Mst, \must,
concentration index and \nuvr\ colour bins, the ``median''
galaxy is a non-detection.
The values of weighted average and median gas fractions illustrated in
this figure are listed in Table~\ref{t_avgs} for reference.
Lastly, small gray circles in these panels represent galaxies in the
GASS parent sample detected by ALFALFA. It is clear that the
shallower, blind \hi\ survey is biased to significantly higher gas
fractions compared to our estimates of the global average. 

As the plots in Figure~\ref{scalings} and Table~\ref{t_avgs} show, 
the gas content of massive galaxies decreases with increasing \Mst, \must\ 
concentration index, and observed \nuvr\ colour.
The strongest correlations are with \must\ and \nuvr.
The average gas fraction decreases by more than a factor of 30
as \must\ increases from $10^8$ \Msun~kpc$^{-2}$ to 
a few times $10^9$ \Msun~kpc$^{-2}$, \ie\ the relation is
very close to a linear one. A similar large decrease is 
obtained as a function of \nuvr. In contrast, the 
average \hi\ fraction decreases by a factor of \about 20
over a 1.5 dex range in stellar mass. The relation with
concentration index is even shallower, remaining approximately
constant up to a concentration index of 2.5, and then declining
by a factor of only \about 5 up to the highest values of 
$R_{90}/R_{50}$.  
Figure~\ref{scalings} also shows that the difference between the mean
and median values of \Mhi/\Mst\ is smallest when it is plotted as a function
of \must\ and \nuvr. This is because these two properties 
yield relatively tight correlations without significant tails to low values
of gas mass fraction (see Fig.~\ref{dr1gf}).  

In Figure~\ref{split}, we split our sample into two  bins in  stellar
mass density (left) and in stellar mass (right).  Although somewhat
limited by small number statistics, there is evidence that a change in
\Mst\ has a smaller effect on the relation between gas fraction and
\must\ than viceversa. In other words, the gas mass fraction of
massive galaxies appears to be primarily correlated with \must, and not \Mst. 

Finally, in Figure~\ref{frac}, we have plotted the fraction of galaxies with
\hi\ gas fractions greater than 0.1 (squares) and greater than 0.03 (circles)
as a function of stellar mass and stellar surface density. As can be seen, 
the fraction of galaxies with significant (\ie, more than a few percent)
gas decreases smoothly as a function of stellar mass.  In contrast, there appears
to be a much more sudden drop in the fraction of such galaxies above
a characteristic stellar surface density of $3 \times 10^8$ \Msun~kpc$^{-2}$.
This result can also be seen in the top right panel of Figure~\ref{dr1gf}, where we see that
{\em all} the galaxies that were not detected in \hi\ have     
stellar surface densities greater than  $3 \times 10^8$ \Msun~kpc$^{-2}$.

\begin{figure*}
\includegraphics[width=14cm]{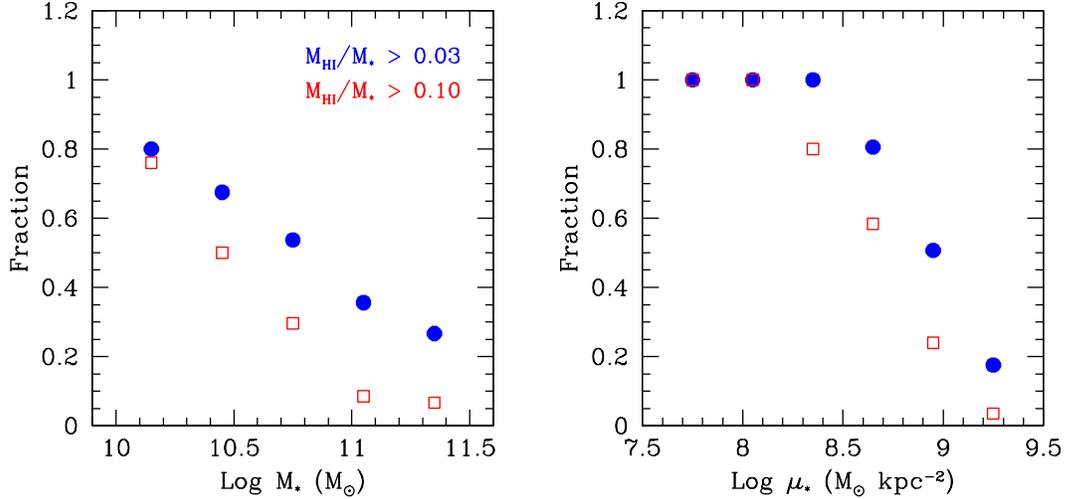}
\caption{Detection fraction of galaxies as a function of stellar mass
(left) and stellar mass surface density (right). Circles and squares
represent objects with \Mhi/\Mst\ greater that 3\% and 10\%, respectively.}
\label{frac}
\end{figure*}

\subsection{Predicting the \hi\ Fraction}\label{s_plane}

As discussed in the previous sections, the \hi\ mass fraction is 
most tightly correlated with stellar surface mass density
and \nuvr\ colour,  and the correlations are
close to linear (at least for galaxies on the blue
sequence). In addition, it can be demonstrated that
there is also a tight correlation  between \nuvr\ colour and
stellar surface mass density. This implies that a
linear combination of the two latter quantities
might provide us with an excellent predictor for the average gas
content of a massive galaxy. We have thus fit a plane to the 2-dimensional
relation between \hi\ mass fraction, stellar surface mass density, and
\nuvr\ colour following the methodology outlined in
\citet{bernardi03}. This is standard practice for elliptical galaxies,
which have been shown to obey a tight plane in the 3-dimensional space
of effective radius, surface brightness and stellar velocity
dispersion. 

We performed  this plane-fitting exercise using only detected galaxies
in one of the non-bootstrapped catalogs described in \S~\ref{s_repr}
(the one illustrated in Figure~\ref{dr1gf}).
The sample used is reproduced in Figure~\ref{plane}, where red and
blue circles indicate DR1 detections and \hi-rich objects added to
the sample, respectively. ALFALFA galaxies (small circles) and
non-detections (upside-down triangles) are not used in the
fit and are shown for comparison only.
The gas fractions obtained from our best fit relation are compared
with measured ones in the figure. The 1:1 relation is
indicated by a dotted line, and the values of the
fit coefficients are given in the caption. The rms scatter
in Log \Mhi/\Mst\ about this relation is 0.315 dex, \ie,
we obtain a 4\% and 13\% reduction in the scatter compared to the 1-d
relations involving \nuvr\ and \must, respectively.

\begin{figure*}
\includegraphics[width=10cm]{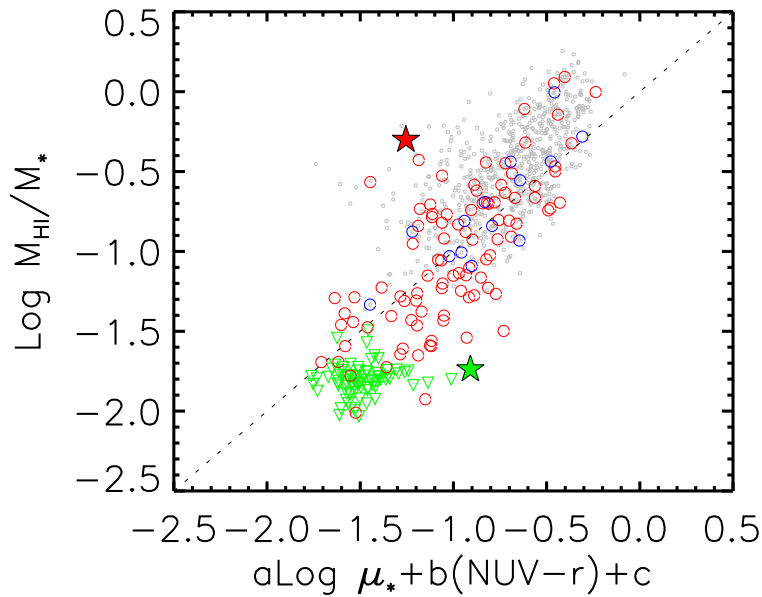}
\caption{The best fit ``plane'' describing the relation between \hi\ mass fraction, stellar mass
surface density and observed \nuvr\ colour is
plotted as a dotted line. All the symbols are the same as those in Fig.~\ref{dr1gf}.
The values of the coefficients listed above are the following: 
$a = -0.332, ~b = -0.240, ~c = 2.856$.}
\label{plane}
\end{figure*}

We note that the most gas-rich galaxies in our sample, as well
as the ALFALFA galaxies, lie systematically above the plane, as expected.  
The most gas-poor galaxies lie systematically below the plane. This may
result from the fact that the non-detections are not used in the fit.
With the increased sample sizes that will be available to us in future,
we will be able to stack the non-detections to estimate an average
gas content, and attempt to use this measurement as a way to anchor
the prediction at the low gas fraction end.

Perhaps the most interesting use for our best-fit plane is as a means                     
to identify interesting objects that {\em deviate}
strongly from the average behavior of the sample. These outliers are
the best candidates for galaxies that might be transitioning between
the blue cloud of star-forming spirals and the red sequence of
passively-evolving galaxies.

Galaxies which are anomalously gas-rich given
their colours and densities scatter above the mean
relation, while those that are gas-poor scatter below. This is clearly
demonstrated by the \hi-rich ALFALFA galaxies, which are preferentially
found above the line. Also, GASS 3505 (marked
with a red star on the diagram), a galaxy that has optical morphology
and colours characteristic of a normal elliptical, but a 50\% \hi\ mass
fraction, is a clear outlier in this plane. 

Of equal interest are the galaxies with low \hi\ mass fractions, but
that are still forming stars. These galaxies are found near the bottom
the plots, but shifted to the right (\eg, GASS 7050, a gas-poor disk
galaxy that was not detected in \hi, is indicated by a green star). 
These may be systems where the
\hi\ gas has recently been stripped by tidal interactions or by
ram-pressure exerted by intergalactic gas, or where other feedback
processes have expelled the gas. 
In future work, we plan to investigate these different classes of
transition galaxy in more detail.

\section{Summary and Discussion}\label{s_disc}

In this  paper we  introduce the GALEX Arecibo SDSS Survey
(GASS), an on-going large program that is gathering high quality \hi-line
spectra for an unbiased sample of \about 1000 massive galaxies using
the Arecibo radio telescope. The sources are selected
from the SDSS spectroscopic and GALEX imaging surveys, and have stellar masses 
\Mst $>10^{10}$ \Msun\ and redshifts $0.025 < z < 0.05$. They are observed 
until detected or until a low gas mass fraction limit (1.5$-$5\% depending on
stellar mass) is reached. 

Blind \hi\ surveys such as ALFALFA, which
detects \about 20\% of the GASS targets, are heavily biased towards
blue, gas-rich systems at the same redshifts. A number of past studies have
investigated the \hi\ content of elliptical and early-type
galaxies \citep[\eg,][]{mor06,oos07}, but these samples are selected by
morphology and hence may not be representative of the full population
of massive galaxies. GASS is the first study to specifically target a 
sample that is homogeneously selected by stellar mass, a robust
measure that is essential for understanding the observations and
connecting them with theory. 

We have presented the first Data Release, DR1, consisting of  \about
20\% of the final GASS sample. Based on this first data installment
we have built an unbiased, representative sample, which has been used  
to explore the main scaling relations between \hi\ gas fraction
and other parameters related to structure and stellar populations of
the galaxies in this study.
Our main findings are discussed below.
\begin{enumerate}
\item A large fraction (\about 60\%) of the galaxies are detected in \hi. Even at
stellar masses above $10^{11}$ \Msun, the detected fraction does not fall
below \about 40\%. Around 9\% of galaxies more massive than $10^{11}$ \Msun\
have \hi\ fractions larger than 0.1 and 34\% have \hi\ fractions larger than 0.03.


\item We have studied the correlation between
\Mhi/\Mst\ ratio and a variety of different galaxy 
properties.  We find that the gas fraction of massive galaxies correlates
strongly with stellar mass, stellar surface mass density and
\nuvr\ colour, but correlates only weakly with the concentration 
index of the \rband\ light. The scatter in the  correlations  increases visibly outside the
blue sequence of \hi-rich, star-forming spirals. 

\item The gas content of massive galaxies decreases for increasing
values of \Mst\ and \must. The latter quantities are clearly related,
but we presented evidence (although based on somewhat limited
statistics) that the primary correlation is with stellar mass surface
density, rather than with stellar mass.

\item  We also found that the fraction of galaxies with significant gas
content (\ie, \Mhi/\Mst\ greater than a few percent) decreases strongly 
above a stellar surface mass density of  $10^{8.5}$ \Msun~kpc$^{-2}$. 
This is the threshold stellar mass
density between disk-dominated, late-type galaxies and bulge-dominated,
early-type objects identified by \citet{kau06}, across which the
recent star formation histories of local galaxies have been
shown to undergo a transition. Based on their study of the scatter in
colour and spectral properties of SDSS galaxies,
Kauffmann et al. argue that star formation proceeds at the same
average rate per unit stellar mass 
below the characteristic surface density, and shuts down
above the threshold. 
\citet{david07} also noted a striking change in the UV-derived
specific SFRs of SDSS galaxies  at the same
characteristic \must\ (see their Fig. 20). 

\item  A similar transition in gas properties near the characteristic
stellar mass \Mst \about 3 \x $10^{10}$ \Msun\ 
\citep{str01,kau03b,bal04} is not evident from our data. 
This appears to be in contradiction to the results
of \citet{bothwell09}, who claim to detect such a
transition in \Mhi/\Mst, based on a sample 
of much more nearby galaxies.
We caution that our results are still
limited by small statistics, particularly at the
low stellar mass end. 
\end {enumerate}

One of the key goals of the GASS survey is to identify and
quantify the incidence of {\em transition} objects, which might be
moving between the blue, star-forming cloud and the red sequence of
passively-evolving galaxies. Depending on their path to or from the
red sequence, these objects should show signs of recent quenching
of star formation or
accretion of gas, respectively. This task requires us to establish 
the {\em normal} gas content of a galaxy of given
mass, structural properties and star formation rate. The classic
concept of \hi\ {\em deficiency} established for spiral galaxies 
\citep{hg84,sgh96} is a well-known attempt to quantify whether
or not a galaxy has been recently stripped of its \hi\ gas.
Our approach is to fit a
plane to the 2-dimensional relation between \hi\ mass fraction,
stellar surface mass density, and \nuvr\ colour. We
showed preliminary results that make use of \hi\ detections only, and
will include non-detections at a later stage, when we have larger samples that
will allow us to recover a signal from their stacked spectra.

The GASS survey has already identified a few interesting examples    
of transition galaxies. In this work we have pointed out two examples,
GASS 3505 and GASS 7050, \ie, a very gas-rich, early-type galaxy and a
gas-poor disk. Many examples of \hi-deficient galaxies are known
from the literature. More intriguing are red sequence
galaxies that might be accreting gas and maybe even regrowing a
disk (such as perhaps GASS 3505). Two such systems, NGC 4203
and NGC 4262, have been identified by \citet{luca09} from a more local
sample of galaxies. The authors report convincing evidence that the \hi, which
is distributed in ring-like structures in both NGC objects, might be of
external origin. Interestingly, also GASS 3505 shows an external ring-like
structure in the GALEX images. We have recently obtained
VLA data for this galaxy, and will report on it in a separate
paper.\\

The GASS data base of stellar and gas-dynamical measurements will 
provide an unprecedented view of the gas properties and kinematics of
massive galaxies, complementing the results of on-going blind surveys
such as ALFALFA. Once completed, GASS will allow us to move beyond the
mean gas fraction scaling relations studied in this paper, and address
second order questions, such as how the gas fractions depend on
metallicity or other quantities at fixed stellar mass. GASS will also
quantify the frequency with which different kinds of transition
galaxies occur in the local universe, and how this depends on factors
such as local environment and AGN content. Insight into the 
nature of massive and transition objects
will give us a strong foundation to further understand the
gas properties of high-redshift galaxies, and will help 
guide future directions for \hi\ surveys at existing and planned radio
facilities.

\section*{Acknowledgments}

B.C. wishes to thank the Arecibo staff, in particular Phil Perillat,
Ganesan Rajagopalan and the telescope operators for their assistance,
and Hector Hernandez for scheduling the observations.  B.C. also
thanks Roderik Overzier and Luca Cortese for helpful comments on the manuscript.

R.G. and M.P.H. acknowledge support from NSF grant AST-0607007 and
from the Brinson Foundation.

The Arecibo Observatory is part of the National Astronomy and
Ionosphere Center, which is operated by Cornell University under a
cooperative agreement with the National Science Foundation. 

GALEX (Galaxy Evolution Explorer) is a NASA Small Explorer, launched
in April 2003. We gratefully acknowledge NASA's support for
construction, operation, and science analysis for the GALEX mission,
developed in cooperation with the Centre National d'Etudes Spatiales
(CNES) of France and the Korean Ministry of Science and Technology.

Funding for the SDSS and SDSS-II has been provided by the Alfred
P. Sloan Foundation, the Participating Institutions, the National
Science Foundation, the U.S. Department of Energy, the National
Aeronautics and Space Administration, the Japanese Monbukagakusho, the
Max Planck Society, and the Higher Education Funding Council for
England. The SDSS Web Site is http://www.sdss.org/.

The SDSS is managed by the Astrophysical Research Consortium for the
Participating Institutions. The Participating Institutions are the
American Museum of Natural History, Astrophysical Institute Potsdam,
University of Basel, University of Cambridge, Case Western Reserve
University, University of Chicago, Drexel University, Fermilab, the
Institute for Advanced Study, the Japan Participation Group, Johns
Hopkins University, the Joint Institute for Nuclear Astrophysics, the
Kavli Institute for Particle Astrophysics and Cosmology, the Korean
Scientist Group, the Chinese Academy of Sciences (LAMOST), Los Alamos
National Laboratory, the Max-Planck-Institute for Astronomy (MPIA),
the Max-Planck-Institute for Astrophysics (MPA), New Mexico State
University, Ohio State University, University of Pittsburgh,
University of Portsmouth, Princeton University, the United States
Naval Observatory, and the University of Washington.

\section*{Appendix: Notes on Individual Objects}\label{s_notes}

We list here notes for galaxies marked with an asterisk in
the last column of Tables~\ref{t_det} and \ref{t_ndet}.
The galaxies are ordered by increasing GASS number. In what follows, 
AA1 and AA2 are abbreviations for ALFALFA detection codes 1 and 2, respectively.\\

\noindent
{\bf Detections (Table \ref{t_det})}\\
{\bf 3261} -- detected on top of weak RFI across all band; AA1.\\
{\bf 3504} -- detected on top of weak RFI across all band; low frequency edge uncertain, systematic error.\\
{\bf 3645} -- detected on top of weak RFI across all band; AA2.\\
{\bf 3777} -- detected on top of RFI across all band; AA2.\\
{\bf 3817} -- hints of RFI across band; AA1.\\
{\bf 3962} -- disk \about 1.5\arcmin\ N has z=0.027, no contamination problems; blue spiral \about 3\arcmin\ E, z=0.0314 (1377.16 MHz), detected in board 4. Low frequency edge uncertain, systematic error.\\
{\bf 3971} -- hints of RFI across band in one pair; AA2; low frequency edge uncertain, systematic error.\\
{\bf 6375} -- group: 2 blue galaxies 2.7\arcmin\ away, GASS 6373 (= VCC 1016, spiral, similar size) to the S and SDSS J122716.82+031812.9 (smaller) to the W, same cz. The two spirals do not look distorted and the \hi\ profile is not clearly confused, just slightly offset in z (but notice that GASS 6373 is offset in the wrong direction). Contamination likely.\\        
{\bf 6506} -- GASS 6501, red galaxy \about 3\arcmin\ away, 1355.75 MHz not detected; RFI spike at 1350 MHz.\\
{\bf 7025} -- high frequency peak uncertain.\\
{\bf 7493} -- GASS 7457 (non-detection) \about 1\arcmin\ away, no contamination problems (z=0.026; GASS 7457 has z=0.036); uncertain profile, systematic error; SDSS emission line z (0.02622) in better agreement with \hi.\\
{\bf 7509} -- small satellites, some contamination very likely; stronger in polarization A; high frequency edge uncertain, systematic error.\\
{\bf 9343} -- 2 blue galaxies within 1\arcmin, no cz, possible contamination; poor fits to edges.\\
{\bf 9463} -- merger, plus another blue galaxy \about 1\arcmin\ away with same z: confusion certain; very uncertain \hi\ profile, systematic error.\\
{\bf 9514} -- smaller disk galaxy \about 1\arcmin\ away, same z, some contamination likely; AA2; poor fit to low frequency edge.\\
{\bf 9619} -- blue, disrupted little galaxy \about 3\arcmin\ away, same z, some contamination possible.\\
{\bf 9776} -- small blue disk \about 15\arcsec\ away, same cz (z=0.027568, 1382.30 MHz), some contamination certain. Looks like the edges of the \hi\ profile come from 9776, but the central peak is dominated by the companion.\\
{\bf 9814} -- blue smudges nearby + small galaxy \about 1.5\arcmin\ NW, same cz; possible contamination; low frequency peak uncertain.\\       
{\bf 10019} -- RFI spike near 1375.3 MHz, slightly drifting in frequency.\\
{\bf 10132} -- two small companions: a blue one \about 1.3\arcmin\ N, strong \Ha\ emission in SDSS spectrum, z=0.0438 (1360.80 MHz), and a galaxy \about 1.5\arcmin\ E, no \Ha\  in SDSS spectrum, z=0.0444 (GASS 10132 has z=0.0443). The strong peak in the \hi\ spectrum is centered on the blue companion. Blend.\\
{\bf 11016} -- galaxy pair: 11016 is a red gal, no emission lines in SDSS spectrum, centered at 1369.07 MHz, the companion is a blue galaxy 1.2\arcmin\ SE, strong emission lines in SDSS spectrum, centered at 1368.93 MHz. Confusion certain.\\
{\bf 11223} -- small early type \about 40\arcsec\ away, no cz, small contamination possible; poor fit to high frequency edge; AA2.\\
{\bf 11386} -- small peak outside \hi\ profile (on the high frequency side) is mostly in polarization B.\\
{\bf 11956} -- companion also detected (SDSS J000814.67+150752.9, z=0.0372, 1369.46 MHz, blue disk 2.1\arcmin\ SW), no overlap.\\
{\bf 11989} -- small blue galaxy 2\arcmin\ away, no cz, but confusion unlikely. High frequency edge uncertain, systematic error; AA4 (i.e., classified by ALFALFA as tentative detection with optical counterpart of unknown redshift).\\
{\bf 12371} -- 3 blue galaxies within 3\arcmin, no contamination problems (largest one has z=0.008, other two have z=0.027 and are detected in board 4 (1383 MHz).\\
{\bf 12983} -- low frequency edge uncertain, systematic error.\\
{\bf 13227} -- stronger in polarization B; uncertain profile, poor fit to high frequency edge, systematic error.\\
{\bf 14831} -- small galaxies nearby (within \about 1\arcmin), no cz; some contamination possible; high frequency edge uncertain, systematic error.\\
{\bf 15181} -- AA1.\\
{\bf 17640} -- AA2.\\
{\bf 18279} -- confused or blend: most of the signal comes from blue galaxy \about 1.5\arcmin\ away, z=0.0492 (1353.80 MHz); 2 other galaxies within 2\arcmin\ have very different z (0.18 and 0.07).\\
{\bf 18421} -- stronger in polarization B.\\
{\bf 18581} -- low frequency peak uncertain.\\
{\bf 20133} -- galaxy pair: GASS 20165 \about 1\arcmin\ away, z=0.0498 (1353.04 MHz; GASS 20133 has z=0.0489, 1354.19 MHz); however 20165 is a red, early type gal, not detected in 10m on-source, so significant contamination is unlikely.\\
{\bf 20144} -- uncertain profile edges, systematic error.\\
{\bf 20286} -- RFI at 1376 MHz.\\
{\bf 23315} -- blue disk \about 1\arcmin\ away has z=0.054, no contamination; low frequency edge uncertain, systematic error.\\
{\bf 23445} -- 2 galaxies at \about 2.7\arcmin: one at N has z=0.08, one at W is small, has \about same cz (z=0.047065, 1356.56 MHz) and is not detected in spectrum (a bit offset so should be visible if there); better in polarization B; uncertain profile.\\
{\bf 24740} -- uncertain profile, systematic error.\\
{\bf 26822} -- AA2.\\
{\bf 28526} -- RFI spike near 1354 MHz (within profile), 2 channels replaced by interpolation.\\
{\bf 29487} -- small companion 2.4\arcmin\ S, z=0.0424 (1362.63 MHz, 0.5 MHz offset), contamination possible but unlikely; strong, narrow RFI near 1359 MHz.\\
{\bf 29505} -- much stronger in polarization B; no trace of RFI, well centered on SDSS z.\\
{\bf 29842} -- AA2.\\
{\bf 30175} -- blue companion 1.6\arcmin\ away, slightly lower z (z=0.0223, 1389.42 MHz) also detected. Two other galaxies \about 3\arcmin\ away have z=0.08.\\
{\bf 30401} -- AA2.\\
{\bf 31156} -- AA2.\\
{\bf 38462} -- poor fits to edges.\\
{\bf 38703} -- stronger in polarization B.\\
{\bf 38717} -- narrow RFI at 1360 and 1370 MHz (no radar blanker).\\
{\bf 38758} -- small, blue companion \about 0.5\arcmin\ away, no cz; galaxy clearly distorted, contamination very likely.\\
{\bf 38964} -- small satellite to the S? blue, edge-on galaxy $\gtrsim$ 3\arcmin\ away, z=0.033468 (1.6 MHz away), not detected but might be responsible for asymmetry (raising low frequency peak).\\
{\bf 39567} -- GASS 39600 \about 2.5\arcmin\ away to the N, no contamination problems (z=0.044, GASS 39567 has z=0.031); much stronger in polarization B.\\
{\bf 39595} -- AA2. Low frequency edge uncertain, systematic error.\\
{\bf 40007} -- galaxy pair (separation 1.3\arcmin, difference of recessional velocities is 60 \kms); 2 blue spirals, same cz, \about 3\arcmin\ and 4.5\arcmin\ away (group); \hi\ spectrum does not look confused but contamination certain (note z off with respect to SDSS redshift of both galaxies).\\
{\bf 40024} -- group: 3 small galaxies around (1\arcmin-3), 2 with \about same cz, one without cz; some contamination likely; high frequency edge and peak uncertain, systematic error, poor fit.\\
{\bf 40393} -- uncertain peaks.\\
{\bf 40494} -- RFI spike near 1357 MHz (within profile), 3 channels replaced by interpolation.\\
{\bf 40781} -- strong RFI spike at 1356 MHz (within profile), 4 channels replaced by interpolation.\\
{\bf 41969} -- AA1.\\
{\bf 41970} -- small disk galaxy 10\arcsec\ to the W, contamination possible (perhaps higher z, though).\\
{\bf 42015} -- AA2.\\
{\bf 42167} -- 3 disks within 2\arcmin-4\arcmin\ have different z (largest one at \about 4\arcmin, 1373.17 MHz, not detected), no contamination problems.\\
{\bf 47221} -- low frequency edge uncertain, systematic error (little side peak is in polarization B only).\\
{\bf 47405} -- high frequency edge uncertain, systematic error.\\

\noindent
{\bf Non-detections (Table \ref{t_ndet})}\\
{\bf 7286} -- small companion \about 1.5\arcmin\ away, same z.\\
{\bf 7457} -- see GASS 7493 (detection).\\
{\bf 9507} -- RFI spike at 1352 MHz.\\
{\bf 9891} -- blue disk \about 2.5\arcmin\ away, z=0.0243 (1386.71 MHz) not detected.\\
{\bf 10150} -- galaxy \about 1\arcmin\ away has z=0.092, no contamination problems.\\
{\bf 10358} -- group (3 galaxies \about same cz within 3\arcmin).\\
{\bf 10367} -- group (3 galaxies \about same cz within 1.5\arcmin-4\arcmin).\\
{\bf 10404} -- small blue galaxy 20\arcsec\ E, no cz, responsible for small spike at 1369.3 MHz?.\\
{\bf 12455} -- group: 2 other galaxies at 2.5\arcmin\ W and 3.5\arcmin\ E, \about same cz; marginally detected the one at 2.5\arcmin\ (small edge-on disk, z=0.0484, 1354.83 MHz)?.\\
{\bf 12458} -- RFI spike at 1370.2 MHz.\\
{\bf 12460} -- RFI spike at 1350.8 MHz. Detected blue companion (irregular galaxy \about 1\arcmin\ away, no cz) at 1352.2 MHz? Not seen RFI there, but much stronger in polarization B.\\
{\bf 13156} -- AA2.\\
{\bf 13549} -- group: several small gals, \about same z, within 3\arcmin; also a larger early type, no cz.\\
{\bf 18422} -- small galaxy \about 1.3\arcmin\ W, no cz (+ other, smaller ones).\\
{\bf 21023} -- RFI at 1358-59 MHz, slightly drifting in frequency.\\
{\bf 25154} -- galaxy pair, companion is a spiral \about 1\arcmin\ away, optical em lines, z=0.0364 (1370.52 MHz), also not detected.\\
{\bf 25214} -- group (4 galaxies \about same cz within 3\arcmin).\\
{\bf 25575} -- galaxy pair (galaxy \about same cz 20\arcsec\ W).\\
{\bf 26958} -- disturbed morphology, merger? marginally detected companion (blue disk \about 0.5\arcmin\ away, no cz)?.\\
{\bf 29090} -- detected blue edge-on disk \about 3\arcmin\ S; 3 other disks \about 3\arcmin-4\arcmin\ away.\\
{\bf 29420} -- group: two small galaxies \about 2\arcmin\ away, same cz, + others, no cz; looks like a galaxy is detected in the off.\\
{\bf 29699} -- group: 2 galaxies \about same z within 2\arcmin, and another one at z=0.044.\\
{\bf 30479} -- disk \about 1.5\arcmin\ away has z=0.08, no contamination problems.\\
{\bf 38472} -- blue galaxy \about 1\arcmin\ SW has z=0.202, no contamination problems.\\
{\bf 39448} -- blue galaxy \about 1\arcmin\ away, same cz (z=0.0341; GASS 39448 has z=0.0339), contamination certain.\\
{\bf 39465} -- blue irregular galaxy \about 1\arcmin\ NW detected in board 3; galaxy \about 1\arcmin\ SW has z=0.076; 2 galaxies \about 0.5\arcmin\ SW, no cz, marginally detected?.\\
{\bf 39606} -- \about 3\arcmin\ away from GASS 39607, same cz, also not detected.\\
{\bf 39607} -- see GASS 39606.\\
{\bf 40257} -- group: 3 small galaxies within 2.5\arcmin\ E and a large disk \about 2\arcmin\ NW, \about same cz.\\
{\bf 40317} -- group: 3 galaxies within 2.5\arcmin, \about same cz; marginally detected blue irregular, 2\arcmin\ SW, z=0.0404 (1365.25 MHz)?.\\
{\bf 40686} -- group: 4 galaxies within 2\arcmin, \about same cz, one detected (SDSS J131527.88+095243.7, a bluish, edge-on disk \about 2\arcmin\ away, z=0.0505, 1352.12 MHz).\\
{\bf 40790} -- small companion \about 1\arcmin\ away, z=0.0499 (1352.90 MHz), also not detected.\\
{\bf 41974} -- group in background (3 galaxies at 2\arcmin-3\arcmin\ with z=0.045-0.047), no contamination problems.\\
{\bf 42020} -- blue irregular galaxy nearly attached, no cz; neg. spike near 1373 MHz is not RFI and is in both pols... galaxy in off?.\\
{\bf 42156} -- detected SDSS J151719.24+072828.3, a blue disk \about 1.5\arcmin\ SW, z=0.0367 (1370.16 MHz). Possible contribution from GASS 42164 at the same cz (1370.03 MHz), a larger, face-on, red spiral \about 2\arcmin\ away from GASS 42156. Also, small disk, 2\arcmin\ E, z=0.077. No confusion problems.\\

\clearpage

\begin{table*}
\tiny
\centering
\caption{SDSS and UV Parameters of the GASS Galaxies.}
\label{t_sdss}
\begin{tabular}{ccccccccccccc}
\hline
   &  &  & Log \Mst & \Rinz & $R_{50}$ & $R_{90}$ & Log \must  & ext$_r$ & r &\nuvr & T$_{NUV}$ & \tmax \\
GASS  & SDSS ID & $z_{\rm SDSS}$ & (\Msun) & (\arcsec) & (\arcsec)&(\arcsec) & (\Msun~kpc$^{-2}$)& (mag)& (mag) & (mag)& (sec) & (min) \\
(1)  & (2)  & (3)  & (4)  & (5) & (6) & (7) & (8)  & (9)  &  (10) & (11) & (12) & (13)\\
\hline
 11956 & J000820.76+150921.6 & 0.0395 & 10.09 &  2.99 &  3.10 &  6.68 &  8.48 &  0.16 & 16.28 &  3.04 &  1680 &  90 \\
 12025 & J001934.54+161215.0 & 0.0366 & 10.84 &  3.67 &  3.92 & 11.86 &  9.13 &  0.18 & 14.74 &  5.93 &  4835 &  14 \\
 11989 & J002558.89+135545.8 & 0.0419 & 10.69 &  2.53 &  2.66 &  8.02 &  9.18 &  0.22 & 15.14 &  5.79 &  3344 &  47 \\
  3261 & J005532.61+154632.9 & 0.0375 & 10.08 &  2.84 &  3.03 &  7.70 &  8.57 &  0.26 & 15.49 &  2.63 &  1918 &  72 \\
  3645 & J011501.75+152448.6 & 0.0307 & 10.33 &  3.07 &  3.22 &  8.74 &  8.93 &  0.17 & 15.12 &  3.97 &  1440 &  32 \\
  3505 & J011746.76+131924.5 & 0.0479 & 10.21 &  1.90 &  1.99 &  6.56 &  8.83 &  0.10 & 16.35 &  4.92 &  1647 & 198 \\
  3504 & J011823.44+133728.4 & 0.0380 & 10.16 &  6.53 &  8.53 & 15.68 &  7.91 &  0.11 & 15.34 &  2.85 &  1647 &  76 \\
  3777 & J012316.81+143932.4 & 0.0396 & 10.26 &  2.68 &  2.87 &  6.35 &  8.75 &  0.14 & 15.69 &  3.17 &  1654 &  90 \\
  3817 & J014325.96+135116.8 & 0.0450 & 10.07 &  3.94 &  4.38 & 10.70 &  8.11 &  0.18 & 15.76 &  2.36 &  3361 & 153 \\
  4239 & J015816.23+141747.9 & 0.0261 & 10.80 &  4.64 &  5.14 & 17.23 &  9.18 &  0.15 & 13.92 &  5.58 &  1663 &   4 \\
  3962 & J020359.14+141837.3 & 0.0427 & 10.90 &  4.66 &  5.26 & 15.46 &  8.84 &  0.19 & 14.51 &  3.22 &  1664 &  19 \\
  3971 & J020552.48+142516.2 & 0.0426 & 10.43 &  3.85 &  4.38 & 11.99 &  8.54 &  0.19 & 15.82 &  3.33 &  1645 & 122 \\
 57017 & J092229.28+142743.3 & 0.0323 & 10.54 &  3.04 &  3.95 & 12.16 &  9.10 &  0.09 & 16.22 &  4.00 &  2857 &  33 \\
 20133 & J093236.58+095025.9 & 0.0489 & 10.86 &  4.38 &  4.81 & 11.05 &  8.73 &  0.11 & 14.69 &   ... &   ... &  42 \\
 20144 & J093502.01+095512.4 & 0.0496 & 10.06 &  3.76 &  4.08 &  7.39 &  8.06 &  0.08 & 16.27 &  2.54 &   106 & 229 \\
 20286 & J095439.45+092640.7 & 0.0346 & 10.53 &  3.59 &  3.88 & 10.21 &  8.89 &  0.10 & 15.74 &  4.11 &  3178 &  44 \\
 14831 & J100530.26+054019.4 & 0.0444 & 11.21 &  4.60 &  5.00 & 13.62 &  9.13 &  0.05 & 14.59 &  4.25 &   299 &   5 \\
 14943 & J101600.20+061505.2 & 0.0458 & 11.33 &  4.56 &  4.85 & 16.36 &  9.23 &  0.06 & 13.74 &  6.44 &   330 &   4 \\
 18900 & J102001.61+083053.6 & 0.0453 & 10.92 &  2.72 &  2.62 &  8.28 &  9.28 &  0.08 & 14.63 &  5.39 &  6825 &  23 \\
 22999 & J102316.42+115120.4 & 0.0455 & 10.64 &  4.20 &  4.65 &  9.77 &  8.62 &  0.10 & 15.46 &  3.79 &   112 &  83 \\
 26602 & J103347.41+124358.1 & 0.0325 & 10.74 &  3.41 &  3.59 & 11.52 &  9.20 &  0.08 & 14.56 &  6.40 &   211 &  13 \\
 15181 & J104002.96+060114.0 & 0.0468 & 11.18 &  6.62 &  7.36 & 21.66 &  8.73 &  0.06 & 14.95 &  4.78 &  2583 &   8 \\
 23315 & J104200.74+114648.0 & 0.0329 & 10.45 &  2.87 &  3.17 &  7.91 &  9.05 &  0.08 & 15.05 &  4.34 &  1701 &  42 \\
 29505 & J104858.81+130058.5 & 0.0362 & 10.76 &  5.24 &  5.66 & 13.41 &  8.75 &  0.09 & 14.37 &  3.67 &  1676 &  19 \\
 23408 & J105322.36+111050.4 & 0.0430 & 11.04 &  5.49 &  6.26 & 19.95 &  8.84 &  0.06 & 14.20 &  5.53 &  1572 &  10 \\
 23445 & J105602.37+115219.8 & 0.0478 & 11.00 &  5.72 &  6.14 & 17.85 &  8.66 &  0.05 & 14.68 &  5.22 &   200 &  20 \\
 23450 & J105648.58+120535.7 & 0.0476 & 10.81 &  2.37 &  2.64 &  8.26 &  9.24 &  0.05 & 15.19 &  5.51 &   203 &  47 \\
 17659 & J105807.59+091633.9 & 0.0344 & 11.17 &  5.04 &  5.67 & 17.93 &  9.23 &  0.08 & 13.66 &  5.61 &   333 &   2 \\
 17640 & J105929.94+084233.1 & 0.0349 & 11.04 &  4.59 &  5.37 & 16.54 &  9.18 &  0.09 & 14.51 &  5.61 &   442 &   4 \\
  5442 & J110032.51+020657.8 & 0.0394 & 11.10 &  8.26 & 11.11 & 19.42 &  8.62 &  0.12 & 13.95 &  2.99 &  1501 &   6 \\
 29699 & J110818.34+131327.5 & 0.0340 & 11.01 &  6.07 &  7.30 & 24.48 &  8.92 &  0.05 & 14.11 &  5.15 &   205 &   5 \\
 12371 & J111306.40+051403.0 & 0.0432 & 10.80 &  3.87 &  4.69 & 13.54 &  8.89 &  0.17 & 14.72 &  3.34 &  1687 &  33 \\
 12455 & J112017.79+041913.3 & 0.0491 & 11.45 &  4.58 &  4.80 & 16.44 &  9.28 &  0.12 & 13.84 &  6.26 &  1536 &   3 \\
 12460 & J112048.30+035021.0 & 0.0494 & 10.98 &  3.75 &  4.43 & 13.86 &  8.97 &  0.12 & 15.14 &  5.13 &    96 &  25 \\
 12458 & J112118.26+033953.0 & 0.0394 & 11.04 &  3.50 &  3.71 & 11.89 &  9.30 &  0.12 & 14.18 &  5.77 &  2045 &   7 \\
 29842 & J112131.76+132535.7 & 0.0341 & 10.80 &  5.33 &  5.98 & 15.50 &  8.82 &  0.07 & 14.33 &  3.69 &  3072 &  12 \\
 30175 & J121602.67+141121.8 & 0.0254 & 10.74 &  4.71 &  6.35 & 18.42 &  9.13 &  0.11 & 13.98 &   ... &   ... &   5 \\
 24740 & J121718.89+084332.7 & 0.0492 & 10.20 &  3.36 &  3.61 &  8.63 &  8.30 &  0.05 & 16.76 &  4.12 &  2937 & 222 \\
 24741 & J121750.81+082549.0 & 0.0492 & 10.91 &  3.50 &  4.20 & 12.26 &  8.98 &  0.06 & 14.68 &  2.64 &  3034 &  33 \\
 18335 & J121853.94+100010.1 & 0.0431 & 10.87 &  3.05 &  3.20 & 10.13 &  9.17 &  0.06 & 14.83 &  6.03 &  3155 &  23 \\
 18421 & J122006.47+100429.2 & 0.0434 & 10.60 &  3.40 &  3.63 & 12.81 &  8.80 &  0.06 & 14.97 &  4.10 &  3155 &  82 \\
 18279 & J122023.10+085137.0 & 0.0488 & 10.60 &  2.53 &  2.63 &  7.50 &  8.95 &  0.06 & 15.59 &  3.84 &  3034 & 136 \\
 24094 & J122030.18+112027.3 & 0.0431 & 11.08 &  5.24 &  5.69 & 14.57 &  8.91 &  0.09 & 14.26 &  4.31 &  3280 &   9 \\
 24757 & J122048.14+084214.4 & 0.0494 & 10.04 &  5.04 &  6.41 & 13.37 &  7.78 &  0.06 & 15.94 &  2.12 &  3034 & 226 \\
 18422 & J122123.25+095053.0 & 0.0467 & 10.56 &  2.21 &  2.27 &  6.78 &  9.07 &  0.06 & 15.09 &  5.28 &  3155 & 134 \\
 30338 & J122319.58+141813.4 & 0.0418 & 10.95 &  4.37 &  4.93 & 14.39 &  8.97 &  0.11 & 14.70 &  4.76 &   671 &  14 \\
  6375 & J122727.61+031807.6 & 0.0488 & 11.05 &  4.42 &  6.23 & 18.91 &  8.92 &  0.06 & 14.55 &  2.65 &  5659 &  17 \\
 18469 & J123251.49+084423.9 & 0.0338 & 10.15 &  5.02 &  5.32 &  9.57 &  8.24 &  0.05 & 15.19 &  2.51 &  2449 &  47 \\
 18581 & J123443.54+090016.8 & 0.0430 & 10.57 &  3.07 &  3.26 & 10.47 &  8.87 &  0.05 & 14.65 &  3.14 &  2449 &  92 \\
 30401 & J123445.19+143326.5 & 0.0465 & 10.50 &  3.13 &  4.05 & 11.04 &  8.71 &  0.10 & 15.71 &  3.55 &   128 & 173 \\
 12966 & J123632.24+061010.5 & 0.0395 & 11.18 &  5.84 &  6.30 & 16.21 &  9.00 &  0.05 & 13.96 &  5.25 &   200 &   4 \\
 30479 & J123708.06+142426.9 & 0.0308 & 10.29 &  2.15 &  2.25 &  6.55 &  9.19 &  0.09 & 15.35 &  5.46 &  1819 &  32 \\
 28143 & J123711.40+083929.8 & 0.0283 & 10.30 &  3.65 &  4.20 & 12.09 &  8.83 &  0.05 & 14.99 &  3.11 &  1696 &  23 \\
 30471 & J123753.20+141652.7 & 0.0263 & 10.83 &  5.18 &  5.70 & 16.96 &  9.11 &  0.08 & 13.65 &  5.58 &  1819 &   4 \\
 12983 & J124032.46+052119.9 & 0.0466 & 11.04 &  4.24 &  4.76 & 12.89 &  8.98 &  0.06 & 14.68 &  4.35 &   186 &  15 \\
  6506 & J124309.36+033452.2 & 0.0487 & 10.77 &  5.08 &  5.65 & 11.14 &  8.52 &  0.09 & 14.89 &  3.05 &  1014 &  62 \\
 13037 & J124314.97+040502.0 & 0.0485 & 11.03 &  3.23 &  3.46 & 11.36 &  9.18 &  0.08 & 14.72 &  5.85 &   292 &  18 \\
 40393 & J124723.17+081732.6 & 0.0485 & 10.47 &  3.23 &  3.35 &  8.72 &  8.62 &  0.07 & 15.61 &  3.42 &  2641 & 209 \\
  6565 & J124938.19+024520.2 & 0.0476 & 10.89 &  2.65 &  2.75 &  9.05 &  9.22 &  0.09 & 14.91 &  5.88 &  1157 &  33 \\
  6583 & J125055.79+031149.3 & 0.0483 & 11.20 &  4.24 &  4.64 & 15.08 &  9.12 &  0.08 & 14.15 &  5.59 &  1157 &   8 \\
 26822 & J125129.06+134654.5 & 0.0376 & 11.03 &  5.82 &  7.12 & 17.72 &  8.89 &  0.11 & 13.81 &  3.48 &   221 &   6 \\
 40494 & J125704.49+103714.5 & 0.0463 & 11.16 &  4.83 &  5.23 & 13.20 &  9.00 &  0.06 & 14.20 &  3.86 &   233 &   8 \\
 28526 & J125828.59+112535.3 & 0.0487 & 10.32 &  4.41 &  4.74 &  9.20 &  8.19 &  0.10 & 15.71 &  2.91 &   149 & 211 \\
 40500 & J125911.09+103006.0 & 0.0460 & 10.88 &  3.26 &  3.39 & 11.59 &  9.07 &  0.06 & 14.91 &  5.22 &   149 &  29 \\
 30508 & J125926.22+142030.0 & 0.0471 & 11.02 &  4.62 &  5.06 & 17.12 &  8.88 &  0.07 & 14.60 &  5.25 &   291 &  17 \\
 13227 & J125950.03+050251.2 & 0.0483 & 11.22 &  6.57 &  7.71 & 18.47 &  8.75 &  0.11 & 14.18 &  4.45 &   128 &   7 \\
 13156 & J130446.99+035417.8 & 0.0408 & 11.02 &  5.19 &  5.37 & 14.95 &  8.91 &  0.08 & 14.75 &  6.04 &  3371 &   9 \\
 25154 & J130457.41+120444.6 & 0.0358 & 11.13 &  4.92 &  5.59 & 16.77 &  9.18 &  0.08 & 13.81 &  5.93 &   259 &   3 \\
 40570 & J131104.94+084828.3 & 0.0325 & 11.14 &  6.20 &  6.61 & 18.64 &  9.08 &  0.08 & 13.43 &  5.61 &   244 &   2 \\
 25214 & J131232.81+114344.2 & 0.0311 & 11.16 &  5.91 &  6.39 & 20.36 &  9.18 &  0.07 & 13.46 &  5.92 &   128 &   2 \\
 40686 & J131529.82+095100.6 & 0.0496 & 11.15 &  3.84 &  4.96 & 16.40 &  9.13 &  0.07 & 14.44 &  5.24 &   203 &  11 \\
 40781 & J131934.30+102717.5 & 0.0480 & 11.10 &  3.84 &  4.32 & 13.79 &  9.11 &  0.06 & 14.56 &  4.05 &   203 &  13 \\
 40790 & J131944.64+102145.9 & 0.0489 & 11.19 &  4.53 &  4.72 & 15.15 &  9.03 &  0.06 & 14.27 &  5.85 &   203 &   9 \\
 25347 & J133019.15+113042.5 & 0.0378 & 11.09 &  4.23 &  4.64 & 16.75 &  9.22 &  0.08 & 13.91 &  5.90 &   213 &   5 \\
 40024 & J133338.20+131409.6 & 0.0432 & 11.16 &  4.24 &  4.69 & 15.40 &  9.17 &  0.07 & 14.26 &  4.99 &   217 &   6 \\
 40007 & J133542.39+131951.2 & 0.0428 & 11.07 &  5.95 &  6.48 & 15.61 &  8.80 &  0.07 & 14.12 &  3.07 &   217 &   9 \\
 13549 & J134525.31+034823.8 & 0.0325 & 10.82 &  4.16 &  4.38 & 14.63 &  9.11 &  0.07 & 14.15 &  5.66 &  2971 &   9 \\
  7031 & J134647.18+020712.1 & 0.0331 & 10.60 &  2.50 &  2.82 &  8.91 &  9.31 &  0.07 & 15.39 &  5.11 &  3197 &  28 \\
  7050 & J134909.69+024511.5 & 0.0489 & 10.99 &  3.77 &  3.91 &  8.45 &  8.99 &  0.07 & 17.02 &  3.25 &  3220 &  23 \\
  7058 & J135032.13+031138.9 & 0.0324 & 10.84 &  9.31 &  9.88 & 18.64 &  8.42 &  0.08 & 13.72 &  3.29 &  3220 &   8 \\
  7025 & J135033.80+021355.9 & 0.0422 & 10.72 &  3.48 &  4.07 & 11.11 &  8.93 &  0.08 & 15.34 &  3.32 &  3197 &  43 \\
 40317 & J135533.72+144552.7 & 0.0408 & 11.07 &  3.55 &  3.97 & 13.47 &  9.29 &  0.06 & 14.33 &  5.45 &   196 &   7 \\
 40257 & J135842.23+132722.9 & 0.0393 & 11.13 &  5.08 &  5.60 & 18.03 &  9.07 &  0.07 & 13.94 &  5.72 &   240 &   5 \\
 40247 & J135942.61+124412.5 & 0.0392 & 11.35 &  4.93 &  5.38 & 18.15 &  9.32 &  0.09 & 13.60 &  5.69 &   240 &   2 \\
  9301 & J140316.98+042147.4 & 0.0462 & 11.33 &  2.42 &  2.57 &  7.50 &  9.77 &  0.08 & 15.38 &  3.36 & 20281 &   4 \\
 25575 & J140606.72+123013.6 & 0.0379 & 11.22 &  4.37 &  4.99 & 15.34 &  9.32 &  0.08 & 14.02 &  5.52 &   220 &   3 \\
  9287 & J140836.94+032522.6 & 0.0404 & 10.94 &  4.45 &  4.90 & 13.11 &  8.97 &  0.08 & 14.50 &  3.94 &  1679 &  13 \\
 38529 & J140946.81+113505.4 & 0.0382 & 11.32 &  6.21 &  7.21 & 21.57 &  9.11 &  0.06 & 13.69 &  5.57 &   110 &   2 \\
  7286 & J141432.05+031124.9 & 0.0269 & 10.73 &  5.29 &  5.49 & 13.65 &  8.97 &  0.09 & 13.54 &  4.56 &  1696 &   6 \\
  9384 & J141509.84+042840.4 & 0.0491 & 10.79 &  4.04 &  4.65 & 12.13 &  8.73 &  0.08 & 15.14 &  3.72 &  2525 &  58 \\
 38462 & J141545.94+102619.8 & 0.0258 & 10.75 &  6.14 &  6.86 & 16.22 &  8.90 &  0.08 & 14.07 &  4.39 &   110 &   5 \\
 38472 & J141608.76+103543.8 & 0.0264 & 10.19 &  3.20 &  3.30 & 10.14 &  8.89 &  0.07 & 14.99 &  5.45 &   208 &  17 \\
  9343 & J141648.38+033748.2 & 0.0493 & 10.72 &  2.18 &  2.54 &  7.95 &  9.20 &  0.09 & 15.52 &  4.10 &  1706 &  80 \\
 38591 & J141740.51+103459.9 & 0.0271 & 10.28 &  3.97 &  4.60 & 15.87 &  8.77 &  0.08 & 14.91 &  6.03 &   110 &  19 \\
 41323 & J141822.46+080551.0 & 0.0440 & 11.02 &  4.30 &  4.18 & 13.55 &  9.00 &  0.09 & 14.39 &  5.53 &   131 &  13 \\
 30811 & J141845.69+055004.7 & 0.0489 & 11.28 &  4.14 &  4.24 & 13.54 &  9.21 &  0.07 & 14.21 &  5.26 &  1732 &   6 \\
  9507 & J142032.84+050638.3 & 0.0496 & 10.73 &  3.29 &  3.43 &  8.58 &  8.84 &  0.09 & 15.44 &  4.85 &  2740 &  81 \\
  9514 & J142209.71+043116.1 & 0.0267 & 10.51 &  4.03 &  4.86 & 14.41 &  9.00 &  0.09 & 14.88 &  4.53 &  2205 &  17 \\
  9463 & J142505.50+031359.3 & 0.0356 & 10.79 &  3.32 &  3.54 & 11.08 &  9.19 &  0.10 & 14.48 &  4.54 &  1756 &  15 \\
  7499 & J142632.10+024506.0 & 0.0394 & 10.76 &  4.12 &  4.52 & 13.08 &  8.88 &  0.09 & 14.63 &  3.26 &  1692 &  27 \\
  7457 & J142713.78+025048.6 & 0.0357 & 10.81 &  3.67 &  3.86 & 12.20 &  9.12 &  0.09 & 14.38 &  5.82 &  1692 &  14 \\
  7493 & J142720.13+025018.1 & 0.0264 & 10.56 &  4.58 &  5.08 & 12.36 &  8.95 &  0.09 & 14.49 &  3.70 &  1692 &  13 \\
\end{tabular}
\end{table*}

\setcounter{table}{0}
\begin{table*}
\tiny
\centering
\caption{-- {\it continued}}
\begin{tabular}{ccccccccccccc}
\hline
   &  &  & Log \Mst & \Rinz & $R_{50}$ & $R_{90}$ & Log \must  & ext$_r$ & r &\nuvr & T$_{NUV}$ & \tmax \\
GASS  & SDSS ID & $z_{\rm SDSS}$ & (\Msun) & (\arcsec) & (\arcsec)&(\arcsec) & (\Msun~kpc$^{-2}$)& (mag)& (mag) & (mag)& (sec) & (min) \\
(1)  & (2)  & (3)  & (4)  & (5) & (6) & (7) & (8)  & (9)  &  (10) & (11) & (12) & (13)\\
\hline
  9619 & J142833.30+031543.1 & 0.0277 & 11.04 &  8.35 &  8.85 & 24.00 &  8.86 &  0.07 & 13.54 &  4.65 &  1692 &   2 \\
  7509 & J143019.95+030529.0 & 0.0312 & 10.51 &  3.56 &  3.62 &  9.64 &  8.97 &  0.09 & 14.71 &  3.89 &  1690 &  32 \\
  9604 & J143052.86+031608.2 & 0.0316 & 10.80 &  4.19 &  4.54 & 12.85 &  9.10 &  0.08 & 14.43 &  3.69 &  1691 &   9 \\
 41482 & J143152.89+071915.1 & 0.0273 & 11.07 &  5.40 &  6.53 & 19.43 &  9.28 &  0.08 & 13.91 &  5.44 &  1684 &   1 \\
  7561 & J143335.68+023819.3 & 0.0283 & 10.59 &  3.90 &  4.36 & 14.30 &  9.05 &  0.10 & 14.59 &  4.02 &  1690 &  15 \\
  9814 & J143348.34+035724.7 & 0.0293 & 10.87 &  6.01 &  7.20 & 17.00 &  8.93 &  0.10 & 13.90 &  4.21 &  1696 &   5 \\
  9776 & J143446.68+032029.7 & 0.0276 & 11.04 &  5.01 &  5.68 & 18.45 &  9.31 &  0.11 & 14.18 &  5.76 &  1696 &   2 \\
  9572 & J143535.74+034121.2 & 0.0280 & 10.66 &  5.44 &  6.18 & 14.75 &  8.84 &  0.11 & 14.61 &  4.09 &  1696 &  10 \\
 29090 & J143810.20+092009.7 & 0.0303 & 11.01 &  5.49 &  5.42 & 16.63 &  9.11 &  0.10 & 13.69 &  5.86 &   362 &   3 \\
  7581 & J143915.52+024340.9 & 0.0279 & 11.04 &  6.19 &  6.41 & 18.85 &  9.12 &  0.09 & 13.45 &  5.75 &  2701 &   2 \\
  9748 & J143917.94+032206.0 & 0.0279 & 10.83 &  8.09 &  9.19 & 23.71 &  8.67 &  0.09 & 13.85 &  5.74 &  1690 &   5 \\
  9917 & J144025.99+033556.0 & 0.0281 & 10.66 &  2.27 &  2.35 &  7.14 &  9.60 &  0.09 & 14.53 &  5.58 &  1690 &  10 \\
  9704 & J144059.30+030813.5 & 0.0265 & 10.95 &  5.79 &  6.33 & 17.22 &  9.12 &  0.10 & 13.77 &  5.37 &  2701 &   2 \\
  9891 & J144225.70+031354.9 & 0.0258 & 10.97 &  5.20 &  5.46 & 15.54 &  9.27 &  0.09 & 13.53 &  5.19 &  1687 &   2 \\
  9863 & J144725.16+032627.9 & 0.0274 & 10.52 &  2.90 &  3.05 & 10.23 &  9.27 &  0.10 & 14.70 &  5.11 &  2190 &  18 \\
 38751 & J144734.58+091351.7 & 0.0290 & 10.20 &  7.16 &  7.85 & 17.00 &  8.12 &  0.09 & 15.25 &  3.02 & 11717 &  25 \\
 38758 & J144743.50+093217.4 & 0.0291 & 10.78 &  2.66 &  2.89 &  9.62 &  9.55 &  0.08 & 14.10 &  5.37 &   221 &   7 \\
 38752 & J144759.39+092534.4 & 0.0288 & 10.30 &  9.12 & 10.21 & 18.65 &  8.01 &  0.09 & 14.58 &  2.86 & 11717 &  24 \\
 29420 & J144858.71+122924.4 & 0.0474 & 11.26 &  4.74 &  5.08 & 16.22 &  9.09 &  0.08 & 14.10 &  5.44 &    87 &   6 \\
 38748 & J144859.36+093017.1 & 0.0467 & 10.90 &  2.32 &  2.38 &  7.77 &  9.37 &  0.07 & 14.97 &  6.06 & 11717 &  28 \\
 38717 & J144929.32+090445.1 & 0.0405 & 10.41 &  4.48 &  5.04 & 12.85 &  8.44 &  0.08 & 15.35 &  3.02 & 11717 & 100 \\
 38716 & J145008.03+090445.8 & 0.0500 & 10.06 &  3.37 &  3.64 &  9.79 &  8.14 &  0.08 & 16.57 &  2.56 & 11717 & 236 \\
 38718 & J145009.95+091241.5 & 0.0291 & 10.05 &  3.13 &  3.07 &  8.29 &  8.68 &  0.08 & 15.25 &  3.07 & 11717 &  25 \\
 38703 & J145029.39+085937.7 & 0.0401 & 10.28 &  3.19 &  3.36 &  9.54 &  8.60 &  0.08 & 15.50 &  2.76 & 11717 &  96 \\
 10031 & J145106.43+045032.6 & 0.0273 & 10.83 &  4.70 &  4.94 & 13.96 &  9.16 &  0.10 & 13.97 &  5.78 &  1693 &   4 \\
 10019 & J145153.39+032147.7 & 0.0308 & 10.68 &  6.32 &  7.19 & 17.82 &  8.65 &  0.12 & 14.20 &  2.95 &  5768 &  14 \\
 10010 & J145233.83+030840.5 & 0.0277 & 10.82 &  5.63 &  6.19 & 17.35 &  8.99 &  0.13 & 14.24 &  4.12 &  5768 &   5 \\
 10040 & J145235.22+043245.1 & 0.0286 & 10.98 &  5.38 &  6.04 & 16.33 &  9.16 &  0.11 & 13.82 &  5.15 &  3652 &   3 \\
 10132 & J145805.70+031745.7 & 0.0443 & 11.05 &  3.60 &  3.74 & 11.59 &  9.18 &  0.13 & 14.59 &  5.80 &  1702 &  11 \\
 10145 & J145806.39+040603.6 & 0.0444 & 11.06 &  4.12 &  4.57 & 15.21 &  9.07 &  0.15 & 14.22 &  4.46 &  2452 &  11 \\
 10150 & J150026.75+041044.4 & 0.0328 & 10.53 &  3.50 &  3.70 &  9.26 &  8.96 &  0.13 & 14.77 &  3.73 &  2298 &  36 \\
 38964 & J150216.35+115503.2 & 0.0322 & 11.27 &  5.74 &  6.86 & 19.22 &  9.28 &  0.12 & 13.56 &  5.41 &    78 &   1 \\
 10218 & J151140.36+034034.2 & 0.0464 & 10.76 &  4.37 &  5.00 & 13.00 &  8.69 &  0.12 & 14.55 &  2.74 &  1666 &  52 \\
 42015 & J151201.77+061309.3 & 0.0462 & 10.18 &  5.02 &  6.46 & 12.80 &  7.99 &  0.10 & 15.53 &  2.38 &  1488 & 171 \\
 10292 & J151322.09+040701.2 & 0.0426 & 10.44 &  2.45 &  2.54 &  6.77 &  8.95 &  0.11 & 15.63 &  5.53 &  1666 & 124 \\
 42025 & J151507.55+070116.5 & 0.0367 & 10.88 &  4.03 &  4.52 & 13.63 &  9.08 &  0.13 & 14.72 &  4.90 &  1605 &  12 \\
 42020 & J151516.46+063918.5 & 0.0352 & 10.51 &  3.82 &  3.87 & 11.24 &  8.80 &  0.09 & 15.20 &  5.75 &  1696 &  52 \\
 42017 & J151524.84+062654.0 & 0.0452 & 10.55 &  1.67 &  1.77 &  5.37 &  9.33 &  0.10 & 15.59 &  5.81 &  1696 & 125 \\
 41969 & J151531.54+062213.3 & 0.0351 & 10.42 &  5.18 &  5.98 & 16.98 &  8.45 &  0.11 & 14.43 &  2.04 &  1696 &  55 \\
 42140 & J151531.97+072829.0 & 0.0457 & 10.98 &  3.84 &  4.04 & 12.65 &  9.02 &  0.09 & 14.57 &  5.99 &   147 &  18 \\
 10367 & J151553.85+030301.1 & 0.0379 & 11.08 &  4.68 &  5.24 & 13.10 &  9.12 &  0.14 & 14.31 &  4.54 &  2527 &   5 \\
 42141 & J151619.14+070944.4 & 0.0360 & 10.97 &  8.57 & 10.40 & 19.16 &  8.53 &  0.10 & 13.95 &  3.49 &  2261 &   7 \\
 41970 & J151630.15+061408.6 & 0.0458 & 10.65 &  3.38 &  3.96 &  9.76 &  8.80 &  0.11 & 16.27 &  3.77 &  1696 &  84 \\
 10358 & J151711.15+032105.7 & 0.0370 & 11.09 &  4.60 &  4.88 & 15.01 &  9.17 &  0.11 & 14.14 &  5.86 &  3103 &   5 \\
 10404 & J151722.96+041248.9 & 0.0361 & 11.02 &  3.60 &  3.83 & 12.72 &  9.34 &  0.12 & 14.04 &  5.87 & 11092 &   5 \\
 42156 & J151724.33+072921.9 & 0.0336 & 10.52 &  3.48 &  3.72 & 11.28 &  8.93 &  0.09 & 15.08 &  5.79 &  2261 &  42 \\
 41974 & J151758.30+064445.4 & 0.0349 & 10.64 &  4.16 &  4.42 & 12.95 &  8.86 &  0.09 & 14.96 &  5.56 &  2261 &  28 \\
 42175 & J151832.42+070720.7 & 0.0456 & 10.88 &  3.49 &  3.68 & 11.39 &  9.01 &  0.10 & 14.75 &  5.71 &  2261 &  28 \\
 10447 & J151840.93+042505.3 & 0.0471 & 10.68 &  3.59 &  3.75 &  9.84 &  8.76 &  0.12 & 15.39 &  5.45 &  1930 &  80 \\
 42167 & J151853.66+073433.3 & 0.0369 & 10.39 &  2.94 &  3.01 &  9.91 &  8.86 &  0.12 & 15.41 &  5.31 &  2261 &  68 \\
 39469 & J151903.38+080819.4 & 0.0338 & 11.12 &  5.79 &  6.87 & 21.16 &  9.09 &  0.09 & 13.88 &  5.27 &   108 &   3 \\
 39467 & J151953.53+080557.2 & 0.0334 & 10.98 &  5.46 &  5.96 & 17.11 &  9.00 &  0.09 & 14.13 &  5.60 &   108 &   5 \\
 39465 & J152028.70+081706.6 & 0.0372 & 10.50 &  2.56 &  2.59 &  8.00 &  9.09 &  0.08 & 15.29 &  5.81 &   108 &  69 \\
 39448 & J152037.22+080305.7 & 0.0338 & 10.09 &  1.77 &  1.97 &  6.10 &  9.08 &  0.09 & 16.10 &  5.42 &   108 &  47 \\
 39605 & J152559.84+094724.5 & 0.0339 & 10.13 &  1.86 &  2.00 &  6.44 &  9.07 &  0.10 & 15.92 &  5.45 &   175 &  48 \\
 39607 & J152706.28+094746.3 & 0.0438 & 10.90 &  2.60 &  2.74 &  8.60 &  9.33 &  0.10 & 14.78 &   ... &   ... &  22 \\
 39595 & J152716.72+100240.2 & 0.0435 & 10.87 &  3.66 &  4.25 & 13.37 &  9.01 &  0.11 & 14.61 &  3.50 &   188 &  24 \\
 39606 & J152716.73+094603.8 & 0.0437 & 10.65 &  3.06 &  3.24 &  9.09 &  8.94 &  0.11 & 15.20 &  5.41 &   188 &  67 \\
 39567 & J152747.42+093729.6 & 0.0312 & 10.57 &  2.85 &  3.07 &  9.55 &  9.22 &  0.10 & 14.93 &  3.84 &   107 &  25 \\
 26980 & J154010.48+060427.8 & 0.0484 & 10.09 &  3.46 &  3.69 &  9.89 &  8.18 &  0.19 & 16.77 &  3.15 & 11120 & 207 \\
 26958 & J154654.33+055328.3 & 0.0419 & 11.26 &  5.30 &  6.23 & 19.22 &  9.10 &  0.15 & 13.52 &  4.71 &   189 &   4 \\
 47221 & J154902.67+175625.5 & 0.0318 & 10.54 &  4.76 &  5.06 & 11.10 &  8.72 &  0.11 & 14.63 &  3.05 &  4899 &  31 \\
 42402 & J155125.21+254539.0 & 0.0460 & 11.03 &  5.37 &  6.14 & 16.67 &  8.78 &  0.19 & 14.47 &  4.18 &   125 &  15 \\
 21023 & J155636.91+272911.9 & 0.0415 & 10.32 &  1.50 &  1.47 &  4.66 &  9.27 &  0.11 & 16.01 &  5.84 & 10176 & 110 \\
 31156 & J155752.01+041544.3 & 0.0258 & 10.68 &  4.74 &  5.56 & 16.52 &  9.05 &  0.21 & 14.09 &  5.76 &  2173 &   7 \\
 29487 & J155754.56+092435.7 & 0.0428 & 11.17 &  3.91 &  4.24 & 13.35 &  9.26 &  0.12 & 14.07 &  5.56 &  1688 &   6 \\
 47405 & J155928.62+201303.1 & 0.0491 & 10.38 &  3.93 &  4.06 & 10.13 &  8.34 &  0.14 & 15.18 &  2.26 &  2609 & 220 \\
 10817 & J220120.93+121148.1 & 0.0291 & 10.61 &  2.81 &  2.87 &  9.50 &  9.33 &  0.24 & 14.55 &  6.34 &   208 &  15 \\
 10872 & J221321.50+132611.3 & 0.0281 & 10.48 &  4.57 &  4.85 & 13.23 &  8.81 &  0.17 & 14.76 &  5.15 &  7104 &  22 \\
 10884 & J221430.63+130444.9 & 0.0257 & 10.47 &  3.83 &  4.26 & 12.50 &  9.04 &  0.19 & 14.84 &  5.97 &  1696 &  15 \\
 11016 & J223619.86+141852.3 & 0.0375 & 10.25 &  2.54 &  2.86 &  9.18 &  8.83 &  0.16 & 15.91 &  4.63 & 31414 &  72 \\
 11223 & J230616.43+135856.3 & 0.0355 & 10.64 &  4.10 &  4.44 & 10.21 &  8.85 &  0.63 & 14.35 &  3.47 &  1687 &  31 \\
 11298 & J231330.39+140350.0 & 0.0394 & 10.10 &  3.82 &  4.56 & 11.92 &  8.29 &  0.18 & 16.60 &  3.88 &  3264 &  89 \\
 11386 & J232611.29+140148.1 & 0.0462 & 10.56 &  2.21 &  2.52 &  8.72 &  9.08 &  0.13 & 15.78 &  4.69 &  2434 & 129 \\
\hline
\end{tabular}
\end{table*}

\clearpage
\begin{table*}
\tiny
\caption{\hi\ Properties of GASS Detections.}
\label{t_det}
\begin{tabular}{cccccccccccccl}
\hline
      &         &               & $T_{\rm on}$ & $\Delta v$ &     & \whi  & \whi$^c$&  $F$      &  rms & &Log \Mhi  &   & \\
GASS  & SDSS ID & $z_{\rm SDSS}$ & (min)       &  (\kms)    & $z$ &  (\kms)& (\kms) & (Jy \kms) & (mJy)& S/N  & (\Msun) & Log \Mhi/\Mst & Q \\
(1)  & (2)  & (3)  & (4)  & (5)  & (6)  & (7)  & (8)  & (9)  &  (10) & (11) & (12) & (13) & (14)\\
\hline
 11956 & J000820.76+150921.6 & 0.0395 & 40 &  21 & 0.039654 & 305$\pm$13 & 274 & 0.22$\pm$0.04 & 0.19 &  9.9 &   9.18 & $-$0.91 &  1* \\
 11989 & J002558.89+135545.8 & 0.0419 & 48 &  21 & 0.041922 & 249$\pm$21 & 219 & 0.16$\pm$0.03 & 0.17 &  9.0 &   9.10 & $-$1.59 &  1* \\
  3261 & J005532.61+154632.9 & 0.0375 &  5 &  10 & 0.037449 & 168$\pm$ 2 & 152 & 1.52$\pm$0.08 & 0.81 & 32.2 &   9.97 & $-$0.11 &  1* \\
  3645 & J011501.75+152448.6 & 0.0307 & 15 &  15 & 0.030745 & 333$\pm$ 4 & 309 & 0.79$\pm$0.07 & 0.43 & 18.2 &   9.51 & $-$0.82 &  1* \\
  3505 & J011746.76+131924.5 & 0.0479 & 10 &  13 & 0.047936 & 202$\pm$ 3 & 180 & 0.79$\pm$0.07 & 0.54 & 20.2 &   9.91 & $-$0.30 &  1  \\
  3504 & J011823.44+133728.4 & 0.0380 &  5 &  10 & 0.038016 &  55$\pm$15 &  43 & 0.72$\pm$0.06 & 0.78 & 27.8 &   9.66 & $-$0.50 &  1* \\
  3777 & J012316.81+143932.4 & 0.0396 & 10 &  16 & 0.039751 & 261$\pm$14 & 236 & 0.52$\pm$0.07 & 0.49 & 11.9 &   9.56 & $-$0.70 &  1* \\
  3817 & J014325.96+135116.8 & 0.0450 &  5 &  13 & 0.045018 & 292$\pm$ 3 & 267 & 1.62$\pm$0.10 & 0.71 & 26.2 &  10.16 &    0.09 &  1* \\
  3962 & J020359.14+141837.3 & 0.0427 &  9 &  21 & 0.042903 & 470$\pm$18 & 430 & 0.67$\pm$0.10 & 0.43 & 10.2 &   9.74 & $-$1.16 &  1* \\
  3971 & J020552.48+142516.2 & 0.0426 &  8 &  16 & 0.042670 & 383$\pm$12 & 353 & 0.68$\pm$0.09 & 0.47 & 13.4 &   9.74 & $-$0.69 &  1* \\
 57017 & J092229.28+142743.3 & 0.0323 &  5 &  13 & 0.032266 & 379$\pm$ 5 & 355 & 1.47$\pm$0.12 & 0.75 & 20.1 &   9.83 & $-$0.71 &  1  \\
 20133 & J093236.58+095025.9 & 0.0489 & 15 &  21 & 0.048977 & 299$\pm$ 7 & 265 & 0.33$\pm$0.06 & 0.29 &  9.8 &   9.54 & $-$1.32 &  1* \\
 20144 & J093502.01+095512.4 & 0.0496 & 15 &  21 & 0.049618 &  94$\pm$21 &  69 & 0.21$\pm$0.04 & 0.40 &  8.5 &   9.37 & $-$0.69 &  1* \\
 20286 & J095439.45+092640.7 & 0.0346 & 10 &  15 & 0.034717 & 414$\pm$ 3 & 385 & 0.57$\pm$0.10 & 0.55 &  8.9 &   9.48 & $-$1.05 &  1* \\
 14831 & J100530.26+054019.4 & 0.0444 &  5 &  21 & 0.044411 & 490$\pm$20 & 449 & 1.03$\pm$0.12 & 0.50 & 12.8 &   9.95 & $-$1.26 &  1* \\
 22999 & J102316.42+115120.4 & 0.0455 & 45 &  21 & 0.045238 & 457$\pm$22 & 417 & 0.38$\pm$0.05 & 0.21 & 12.5 &   9.54 & $-$1.10 &  1  \\
 15181 & J104002.96+060114.0 & 0.0468 &  5 &  10 & 0.046962 & 563$\pm$ 5 & 528 & 2.25$\pm$0.13 & 0.72 & 24.4 &  10.34 & $-$0.84 &  1* \\
 23315 & J104200.74+114648.0 & 0.0329 & 50 &  21 & 0.033036 & 289$\pm$39 & 260 & 0.13$\pm$0.04 & 0.19 &  6.1 &   8.80 & $-$1.65 &  2* \\
 29505 & J104858.81+130058.5 & 0.0362 & 20 &  21 & 0.036278 & 338$\pm$19 & 306 & 0.29$\pm$0.05 & 0.23 & 10.3 &   9.22 & $-$1.54 &  1* \\
 23445 & J105602.37+115219.8 & 0.0478 & 20 &  16 & 0.048003 & 235$\pm$11 & 209 & 0.24$\pm$0.05 & 0.34 &  8.2 &   9.39 & $-$1.61 &  1* \\
 17640 & J105929.94+084233.1 & 0.0349 &  4 &  21 & 0.034891 & 459$\pm$ 2 & 423 & 0.74$\pm$0.14 & 0.62 &  8.1 &   9.60 & $-$1.44 &  1* \\
  5442 & J110032.51+020657.8 & 0.0394 &  5 &  13 & 0.039494 & 355$\pm$ 2 & 329 & 4.32$\pm$0.15 & 0.97 & 46.8 &  10.47 & $-$0.63 &  1  \\
 12371 & J111306.40+051403.0 & 0.0432 &  5 &  21 & 0.043180 & 390$\pm$ 1 & 354 & 0.91$\pm$0.15 & 0.68 & 10.4 &   9.87 & $-$0.93 &  1* \\
 29842 & J112131.76+132535.7 & 0.0341 &  8 &  15 & 0.034137 & 490$\pm$ 2 & 459 & 0.70$\pm$0.10 & 0.48 & 10.7 &   9.55 & $-$1.25 &  1* \\
 30175 & J121602.67+141121.8 & 0.0254 & 15 &  20 & 0.025448 & 275$\pm$ 2 & 249 & 0.19$\pm$0.06 & 0.33 &  5.5 &   8.74 & $-$2.00 &  2* \\
 24740 & J121718.89+084332.7 & 0.0492 & 90 &  21 & 0.049141 & 246$\pm$27 & 214 & 0.08$\pm$0.02 & 0.13 &  6.1 &   8.93 & $-$1.27 &  2* \\
 24741 & J121750.81+082549.0 & 0.0492 &  5 &  16 & 0.049231 & 434$\pm$11 & 399 & 1.21$\pm$0.12 & 0.60 & 16.4 &  10.11 & $-$0.80 &  1  \\
 18335 & J121853.94+100010.1 & 0.0431 & 20 &  21 & 0.043237 & 513$\pm$ 2 & 472 & 0.46$\pm$0.06 & 0.26 & 10.5 &   9.58 & $-$1.29 &  1  \\
 18421 & J122006.47+100429.2 & 0.0434 & 30 &  21 & 0.043233 & 344$\pm$ 9 & 310 & 0.19$\pm$0.05 & 0.24 &  6.6 &   9.20 & $-$1.40 &  1* \\
 18279 & J122023.10+085137.0 & 0.0488 &  5 &  21 & 0.048911 & 362$\pm$11 & 325 & 0.63$\pm$0.11 & 0.54 &  9.5 &   9.83 & $-$0.77 &  5* \\
 24757 & J122048.14+084214.4 & 0.0494 &  5 &  10 & 0.049418 &  70$\pm$ 3 &  57 & 1.02$\pm$0.05 & 0.75 & 35.9 &  10.04 &    0.00 &  1  \\
 30338 & J122319.58+141813.4 & 0.0418 & 14 &  21 & 0.041916 & 543$\pm$13 & 500 & 0.56$\pm$0.08 & 0.31 & 10.3 &   9.64 & $-$1.31 &  1  \\
  6375 & J122727.61+031807.6 & 0.0488 &  5 &  13 & 0.049067 & 501$\pm$ 6 & 465 & 2.75$\pm$0.15 & 0.79 & 27.3 &  10.47 & $-$0.58 &  1* \\
 18469 & J123251.49+084423.9 & 0.0338 & 10 &  13 & 0.033793 & 223$\pm$ 5 & 204 & 0.52$\pm$0.06 & 0.49 & 14.2 &   9.42 & $-$0.73 &  1  \\
 18581 & J123443.54+090016.8 & 0.0430 & 10 &  21 & 0.042996 & 337$\pm$26 & 303 & 0.91$\pm$0.07 & 0.38 & 20.3 &   9.87 & $-$0.70 &  1* \\
 30401 & J123445.19+143326.5 & 0.0465 &  5 &  16 & 0.046472 & 237$\pm$ 7 & 212 & 0.87$\pm$0.10 & 0.67 & 15.1 &   9.92 & $-$0.58 &  1* \\
 28143 & J123711.40+083929.8 & 0.0283 &  5 &  13 & 0.028270 & 349$\pm$ 1 & 328 & 1.16$\pm$0.11 & 0.67 & 18.3 &   9.61 & $-$0.69 &  1  \\
 12983 & J124032.46+052119.9 & 0.0466 & 20 &  21 & 0.046706 & 440$\pm$40 & 400 & 0.47$\pm$0.07 & 0.31 & 10.7 &   9.66 & $-$1.38 &  1* \\
  6506 & J124309.36+033452.2 & 0.0487 & 10 &  16 & 0.048747 & 377$\pm$ 5 & 344 & 0.86$\pm$0.09 & 0.48 & 16.4 &   9.96 & $-$0.81 &  1* \\
 40393 & J124723.17+081732.6 & 0.0485 & 20 &  21 & 0.048690 & 159$\pm$40 & 132 & 0.25$\pm$0.03 & 0.24 & 12.5 &   9.42 & $-$1.05 &  1* \\
 26822 & J125129.06+134654.5 & 0.0376 &  5 &  16 & 0.037499 & 432$\pm$ 4 & 401 & 1.24$\pm$0.11 & 0.57 & 18.1 &   9.88 & $-$1.15 &  1* \\
 40494 & J125704.49+103714.5 & 0.0463 &  8 &  16 & 0.046345 & 350$\pm$ 6 & 320 & 0.95$\pm$0.08 & 0.44 & 20.6 &   9.96 & $-$1.20 &  1* \\
 28526 & J125828.59+112535.3 & 0.0487 &  9 &  16 & 0.048697 & 301$\pm$ 3 & 272 & 0.51$\pm$0.09 & 0.53 &  9.7 &   9.73 & $-$0.59 &  1* \\
 13227 & J125950.03+050251.2 & 0.0483 &  8 &  21 & 0.048353 & 313$\pm$38 & 278 & 0.44$\pm$0.10 & 0.50 &  7.8 &   9.66 & $-$1.56 &  2* \\
 40781 & J131934.30+102717.5 & 0.0480 & 12 &  21 & 0.047963 & 509$\pm$16 & 465 & 0.88$\pm$0.09 & 0.35 & 15.1 &   9.95 & $-$1.15 &  1* \\
 40024 & J133338.20+131409.6 & 0.0432 & 10 &  21 & 0.043220 & 605$\pm$57 & 560 & 1.03$\pm$0.09 & 0.32 & 16.3 &   9.93 & $-$1.23 &  1* \\
 40007 & J133542.39+131951.2 & 0.0428 &  5 &  16 & 0.042516 & 450$\pm$15 & 417 & 1.39$\pm$0.12 & 0.60 & 18.6 &  10.05 & $-$1.02 &  5* \\
  7031 & J134647.18+020712.1 & 0.0331 & 15 &  21 & 0.033150 & 386$\pm$ 4 & 353 & 0.27$\pm$0.07 & 0.34 &  6.3 &   9.12 & $-$1.48 &  1  \\
  7058 & J135032.13+031138.9 & 0.0324 &  9 &  21 & 0.032442 & 168$\pm$ 6 & 143 & 0.47$\pm$0.08 & 0.55 & 10.3 &   9.34 & $-$1.50 &  1  \\
  7025 & J135033.80+021355.9 & 0.0422 & 10 &  16 & 0.042042 & 435$\pm$20 & 402 & 1.22$\pm$0.11 & 0.55 & 18.3 &   9.98 & $-$0.74 &  1* \\
  9301 & J140316.98+042147.4 & 0.0462 &  4 &  21 & 0.046175 & 345$\pm$10 & 309 & 0.78$\pm$0.13 & 0.66 &  9.8 &   9.87 & $-$1.46 &  1  \\
  9287 & J140836.94+032522.6 & 0.0404 &  5 &  21 & 0.040505 & 346$\pm$ 5 & 313 & 1.06$\pm$0.11 & 0.55 & 15.9 &   9.88 & $-$1.06 &  1  \\
  9384 & J141509.84+042840.4 & 0.0491 &  5 &  16 & 0.049154 & 413$\pm$10 & 378 & 0.75$\pm$0.10 & 0.54 & 12.1 &   9.91 & $-$0.88 &  1  \\
 38462 & J141545.94+102619.8 & 0.0258 & 15 &  21 & 0.025861 & 316$\pm$22 & 288 & 0.23$\pm$0.06 & 0.33 &  6.0 &   8.82 & $-$1.93 &  2* \\
  9343 & J141648.38+033748.2 & 0.0493 & 10 &  21 & 0.049461 & 371$\pm$21 & 333 & 0.89$\pm$0.09 & 0.41 & 17.1 &   9.99 & $-$0.73 &  1* \\
  9514 & J142209.71+043116.1 & 0.0267 &  4 &  21 & 0.026698 & 390$\pm$46 & 359 & 1.16$\pm$0.13 & 0.60 & 15.4 &   9.56 & $-$0.95 &  1* \\
  9463 & J142505.50+031359.3 & 0.0356 & 20 &  21 & 0.035398 & 458$\pm$84 & 422 & 0.26$\pm$0.06 & 0.28 &  6.2 &   9.15 & $-$1.64 &  3* \\
  7499 & J142632.10+024506.0 & 0.0394 &  5 &  10 & 0.039401 & 409$\pm$ 2 & 383 & 2.00$\pm$0.13 & 0.82 & 26.4 &  10.14 & $-$0.62 &  1  \\
  7493 & J142720.13+025018.1 & 0.0264 &  5 &  21 & 0.026101 & 266$\pm$45 & 239 & 0.87$\pm$0.12 & 0.57 & 14.6 &   9.41 & $-$1.15 &  1* \\
  9619 & J142833.30+031543.1 & 0.0277 &  4 &  15 & 0.027753 & 568$\pm$ 2 & 538 & 1.59$\pm$0.19 & 0.87 & 11.7 &   9.73 & $-$1.31 &  1* \\
  7509 & J143019.95+030529.0 & 0.0312 & 15 &  21 & 0.031208 & 321$\pm$21 & 291 & 0.28$\pm$0.06 & 0.33 &  7.5 &   9.08 & $-$1.43 &  1* \\
  9604 & J143052.86+031608.2 & 0.0316 &  5 &  13 & 0.031702 & 416$\pm$ 5 & 391 & 1.71$\pm$0.13 & 0.75 & 21.7 &   9.88 & $-$0.92 &  1  \\
  7561 & J143335.68+023819.3 & 0.0283 & 10 &  10 & 0.028303 & 393$\pm$ 2 & 372 & 1.87$\pm$0.09 & 0.61 & 35.1 &   9.82 & $-$0.77 &  1  \\
  9814 & J143348.34+035724.7 & 0.0293 & 10 &  21 & 0.029490 & 280$\pm$10 & 252 & 0.50$\pm$0.09 & 0.48 &  9.7 &   9.28 & $-$1.59 &  1* \\
  9776 & J143446.68+032029.7 & 0.0276 &  4 &  21 & 0.027349 & 345$\pm$ 2 & 316 & 0.68$\pm$0.12 & 0.61 &  9.3 &   9.35 & $-$1.69 &  1* \\
  9572 & J143535.74+034121.2 & 0.0280 & 10 &  15 & 0.027879 & 333$\pm$ 3 & 309 & 0.78$\pm$0.08 & 0.50 & 15.5 &   9.43 & $-$1.23 &  1  \\
  9863 & J144725.16+032627.9 & 0.0274 & 14 &  10 & 0.027389 & 336$\pm$ 2 & 317 & 2.73$\pm$0.07 & 0.50 & 67.2 &   9.95 & $-$0.57 &  1  \\
 38751 & J144734.58+091351.7 & 0.0290 & 10 &  15 & 0.029043 & 334$\pm$ 2 & 309 & 0.95$\pm$0.09 & 0.51 & 18.3 &   9.54 & $-$0.66 &  1  \\
 38758 & J144743.50+093217.4 & 0.0291 &  8 &  21 & 0.029340 & 381$\pm$ 8 & 350 & 0.55$\pm$0.09 & 0.44 & 10.0 &   9.32 & $-$1.46 &  1* \\
 38752 & J144759.39+092534.4 & 0.0288 &  5 &  15 & 0.028793 & 222$\pm$ 3 & 201 & 1.00$\pm$0.10 & 0.69 & 17.6 &   9.56 & $-$0.74 &  1  \\
 38748 & J144859.36+093017.1 & 0.0467 & 50 &  21 & 0.046889 & 510$\pm$ 5 & 466 & 0.17$\pm$0.04 & 0.18 &  5.6 &   9.21 & $-$1.69 &  2  \\
 38717 & J144929.32+090445.1 & 0.0405 & 10 &  13 & 0.040711 & 317$\pm$ 5 & 293 & 0.76$\pm$0.08 & 0.50 & 16.9 &   9.75 & $-$0.66 &  1* \\
 38716 & J145008.03+090445.8 & 0.0500 &  5 &  13 & 0.049965 & 317$\pm$ 3 & 290 & 1.16$\pm$0.10 & 0.65 & 19.6 &  10.11 &    0.05 &  1  \\
 38718 & J145009.95+091241.5 & 0.0291 &  5 &  21 & 0.029103 & 186$\pm$ 5 & 160 & 0.36$\pm$0.07 & 0.50 &  8.2 &   9.13 & $-$0.92 &  1  \\
 38703 & J145029.39+085937.7 & 0.0401 &  5 &  21 & 0.040121 & 225$\pm$12 & 196 & 0.40$\pm$0.08 & 0.51 &  8.1 &   9.45 & $-$0.83 &  1* \\
 10019 & J145153.39+032147.7 & 0.0308 &  5 &  10 & 0.030815 & 354$\pm$ 1 & 334 & 4.11$\pm$0.11 & 0.80 & 61.5 &  10.23 & $-$0.45 &  1* \\
 10010 & J145233.83+030840.5 & 0.0277 &  5 &  13 & 0.027769 & 442$\pm$ 2 & 418 & 3.25$\pm$0.13 & 0.75 & 39.0 &  10.04 & $-$0.78 &  1  \\
 10132 & J145805.70+031745.7 & 0.0443 & 12 &  21 & 0.044397 & 419$\pm$ 7 & 381 & 0.53$\pm$0.08 & 0.35 & 10.9 &   9.66 & $-$1.39 &  5* \\
 10145 & J145806.39+040603.6 & 0.0444 & 10 &  21 & 0.044464 & 300$\pm$ 4 & 267 & 0.49$\pm$0.07 & 0.36 & 12.0 &   9.63 & $-$1.43 &  1  \\
 38964 & J150216.35+115503.2 & 0.0322 &  9 &  21 & 0.032329 & 666$\pm$ 2 & 625 & 0.39$\pm$0.11 & 0.39 &  4.8 &   9.26 & $-$2.01 &  2* \\
 10218 & J151140.36+034034.2 & 0.0464 &  5 &  10 & 0.046345 & 233$\pm$ 2 & 213 & 1.85$\pm$0.09 & 0.81 & 33.3 &  10.25 & $-$0.51 &  1  \\
 42015 & J151201.77+061309.3 & 0.0462 &  5 &  16 & 0.046172 & 178$\pm$11 & 155 & 0.76$\pm$0.12 & 0.92 & 11.1 &   9.86 & $-$0.32 &  1* \\
 42025 & J151507.55+070116.5 & 0.0367 & 12 &  15 & 0.036712 & 409$\pm$ 3 & 379 & 0.49$\pm$0.07 & 0.39 & 11.0 &   9.47 & $-$1.40 &  1  \\
 41969 & J151531.54+062213.3 & 0.0351 &  5 &   5 & 0.035088 & 109$\pm$ 1 & 101 & 3.50$\pm$0.07 & 1.21 & 87.7 &  10.28 & $-$0.14 &  1* \\
 42141 & J151619.14+070944.4 & 0.0360 &  8 &  15 & 0.036055 & 358$\pm$ 4 & 331 & 0.96$\pm$0.09 & 0.54 & 16.8 &   9.74 & $-$1.23 &  1  \\
 41970 & J151630.15+061408.6 & 0.0458 & 25 &  21 & 0.045798 & 438$\pm$ 4 & 398 & 0.35$\pm$0.06 & 0.28 &  8.6 &   9.51 & $-$1.14 &  1* \\
 42167 & J151853.66+073433.3 & 0.0369 & 64 &  21 & 0.036929 & 208$\pm$ 5 & 181 & 0.08$\pm$0.02 & 0.14 &  6.0 &   8.66 & $-$1.73 &  1* \\
 39595 & J152716.72+100240.2 & 0.0435 &  2 &  21 & 0.043447 & 381$\pm$21 & 345 & 1.32$\pm$0.23 & 1.09 &  9.5 &  10.04 & $-$0.83 &  1* \\
 39567 & J152747.42+093729.6 & 0.0312 & 20 &  21 & 0.031252 & 353$\pm$ 7 & 322 & 0.22$\pm$0.06 & 0.28 &  6.5 &   8.98 & $-$1.59 &  1* \\
 26980 & J154010.48+060427.8 & 0.0484 &  9 &  16 & 0.048300 & 307$\pm$ 2 & 278 & 0.57$\pm$0.09 & 0.55 & 10.5 &   9.77 & $-$0.32 &  1  \\
 47221 & J154902.67+175625.5 & 0.0318 & 14 &  21 & 0.031792 & 283$\pm$30 & 255 & 0.42$\pm$0.07 & 0.39 & 10.0 &   9.27 & $-$1.27 &  1* \\
 42402 & J155125.21+254539.0 & 0.0460 &  5 &  13 & 0.046055 & 525$\pm$ 3 & 489 & 3.33$\pm$0.13 & 0.69 & 36.1 &  10.50 & $-$0.53 &  1  \\
 31156 & J155752.01+041544.3 & 0.0258 &  4 &  15 & 0.025791 & 213$\pm$ 3 & 193 & 0.85$\pm$0.10 & 0.72 & 14.6 &   9.39 & $-$1.29 &  1* \\
 29487 & J155754.56+092435.7 & 0.0428 & 10 &  16 & 0.042860 & 232$\pm$ 8 & 207 & 0.30$\pm$0.07 & 0.52 &  6.9 &   9.39 & $-$1.78 &  2* \\
 47405 & J155928.62+201303.1 & 0.0491 & 10 &  16 & 0.049134 & 214$\pm$17 & 189 & 0.76$\pm$0.07 & 0.50 & 18.6 &   9.91 & $-$0.47 &  1* \\
 11016 & J223619.86+141852.3 & 0.0375 &  5 &  16 & 0.037593 & 389$\pm$ 2 & 360 & 1.05$\pm$0.12 & 0.68 & 14.1 &   9.82 & $-$0.43 &  5* \\
 11223 & J230616.43+135856.3 & 0.0355 & 20 &  21 & 0.035611 & 344$\pm$33 & 312 & 0.40$\pm$0.06 & 0.30 & 11.1 &   9.35 & $-$1.29 &  1* \\
 11298 & J231330.39+140350.0 & 0.0394 & 10 &  16 & 0.039444 & 349$\pm$ 2 & 321 & 0.67$\pm$0.08 & 0.48 & 13.4 &   9.66 & $-$0.44 &  1  \\
 11386 & J232611.29+140148.1 & 0.0462 & 43 &  21 & 0.046285 & 336$\pm$ 1 & 300 & 0.20$\pm$0.04 & 0.18 &  9.2 &   9.28 & $-$1.28 &  1* \\
\hline
\end{tabular}
\end{table*}

\clearpage
\begin{table*}
\scriptsize
\centering
\caption{GASS Non-detections.}
\label{t_ndet}
\begin{tabular}{ccccccccc}
\hline
      &         &              &  $T_{\rm on}$ &  rms  & Log \Mhi$_{,lim}$ &  \\
GASS  & SDSS ID & $z_{\rm SDSS}$& (min)        &  (mJy)& (\Msun)  & Log \Mhi$_{,lim}$/\Mst & Note\\
(1)  & (2)  & (3)  & (4)  & (5)  & (6)  & (7) & (8)\\
\hline
 12025 & J001934.54+161215.0 & 0.0366 &  14 &  0.43 & $<\,$9.10  &  $<-$1.74  &  ...  \\
  4239 & J015816.23+141747.9 & 0.0261 &   5 &  0.58 & $<\,$8.92  &  $<-$1.88  &  ...  \\
 14943 & J101600.20+061505.2 & 0.0458 &   4 &  0.69 & $<\,$9.50  &  $<-$1.83  &  ...  \\
 18900 & J102001.61+083053.6 & 0.0453 &  23 &  0.28 & $<\,$9.09  &  $<-$1.83  &  ...  \\
 26602 & J103347.41+124358.1 & 0.0325 &  15 &  0.40 & $<\,$8.95  &  $<-$1.80  &  ...  \\
 23408 & J105322.36+111050.4 & 0.0430 &  10 &  0.45 & $<\,$9.26  &  $<-$1.78  &  ...  \\
 23450 & J105648.58+120535.7 & 0.0476 &  48 &  0.18 & $<\,$8.96  &  $<-$1.85  &  ...  \\
 17659 & J105807.59+091633.9 & 0.0344 &   4 &  0.74 & $<\,$9.27  &  $<-$1.90  &  ...  \\
 29699 & J110818.34+131327.5 & 0.0340 &   5 &  0.66 & $<\,$9.21  &  $<-$1.80  &   *   \\
 12455 & J112017.79+041913.3 & 0.0491 &   4 &  0.72 & $<\,$9.58  &  $<-$1.87  &   *   \\
 12460 & J112048.30+035021.0 & 0.0494 &  25 &  0.29 & $<\,$9.19  &  $<-$1.79  &   *   \\
 12458 & J112118.26+033953.0 & 0.0394 &   8 &  0.47 & $<\,$9.20  &  $<-$1.84  &   *   \\
 24094 & J122030.18+112027.3 & 0.0431 &  10 &  0.46 & $<\,$9.26  &  $<-$1.82  &  ...  \\
 18422 & J122123.25+095053.0 & 0.0467 &  90 &  0.17 & $<\,$8.91  &  $<-$1.65  &   *   \\
 12966 & J123632.24+061010.5 & 0.0395 &   4 &  0.72 & $<\,$9.38  &  $<-$1.80  &  ...  \\
 30479 & J123708.06+142426.9 & 0.0308 &  70 &  0.17 & $<\,$8.54  &  $<-$1.75  &   *   \\
 30471 & J123753.20+141652.7 & 0.0263 &   4 &  0.68 & $<\,$9.00  &  $<-$1.83  &  ...  \\
 13037 & J124314.97+040502.0 & 0.0485 &  18 &  0.35 & $<\,$9.26  &  $<-$1.77  &  ...  \\
  6565 & J124938.19+024520.2 & 0.0476 &  38 &  0.28 & $<\,$9.13  &  $<-$1.76  &  ...  \\
  6583 & J125055.79+031149.3 & 0.0483 &   8 &  0.59 & $<\,$9.48  &  $<-$1.72  &  ...  \\
 40500 & J125911.09+103006.0 & 0.0460 &  29 &  0.28 & $<\,$9.12  &  $<-$1.76  &  ...  \\
 30508 & J125926.22+142030.0 & 0.0471 &  17 &  0.39 & $<\,$9.28  &  $<-$1.74  &  ...  \\
 13156 & J130446.99+035417.8 & 0.0408 &   9 &  0.59 & $<\,$9.32  &  $<-$1.70  &   *   \\
 25154 & J130457.41+120444.6 & 0.0358 &   4 &  0.75 & $<\,$9.32  &  $<-$1.81  &   *   \\
 40570 & J131104.94+084828.3 & 0.0325 &   4 &  0.73 & $<\,$9.22  &  $<-$1.92  &  ...  \\
 25214 & J131232.81+114344.2 & 0.0311 &   4 &  0.68 & $<\,$9.14  &  $<-$2.02  &   *   \\
 40686 & J131529.82+095100.6 & 0.0496 &  10 &  0.39 & $<\,$9.32  &  $<-$1.84  &   *   \\
 40790 & J131944.64+102145.9 & 0.0489 &   9 &  0.44 & $<\,$9.36  &  $<-$1.83  &   *   \\
 25347 & J133019.15+113042.5 & 0.0378 &   5 &  0.59 & $<\,$9.26  &  $<-$1.83  &  ...  \\
 13549 & J134525.31+034823.8 & 0.0325 &  10 &  0.49 & $<\,$9.04  &  $<-$1.78  &   *   \\
  7050 & J134909.69+024511.5 & 0.0489 &  23 &  0.34 & $<\,$9.25  &  $<-$1.74  &  ...  \\
 40317 & J135533.72+144552.7 & 0.0408 &   8 &  0.52 & $<\,$9.27  &  $<-$1.80  &   *   \\
 40257 & J135842.23+132722.9 & 0.0393 &   5 &  0.63 & $<\,$9.32  &  $<-$1.81  &   *   \\
 40247 & J135942.61+124412.5 & 0.0392 &   4 &  0.70 & $<\,$9.37  &  $<-$1.98  &  ...  \\
 25575 & J140606.72+123013.6 & 0.0379 &   5 &  0.60 & $<\,$9.27  &  $<-$1.95  &   *   \\
 38529 & J140946.81+113505.4 & 0.0382 &   4 &  0.63 & $<\,$9.29  &  $<-$2.03  &  ...  \\
  7286 & J141432.05+031124.9 & 0.0269 &  10 &  0.51 & $<\,$8.89  &  $<-$1.84  &   *   \\
 38472 & J141608.76+103543.8 & 0.0264 &  43 &  0.22 & $<\,$8.52  &  $<-$1.67  &   *   \\
 38591 & J141740.51+103459.9 & 0.0271 &  48 &  0.19 & $<\,$8.46  &  $<-$1.82  &  ...  \\
 41323 & J141822.46+080551.0 & 0.0440 &  13 &  0.35 & $<\,$9.16  &  $<-$1.86  &  ...  \\
 30811 & J141845.69+055004.7 & 0.0489 &   5 &  0.65 & $<\,$9.53  &  $<-$1.75  &  ...  \\
  9507 & J142032.84+050638.3 & 0.0496 &  75 &  0.18 & $<\,$8.98  &  $<-$1.75  &   *   \\
  7457 & J142713.78+025048.6 & 0.0357 &  15 &  0.40 & $<\,$9.03  &  $<-$1.78  &   *   \\
 41482 & J143152.89+071915.1 & 0.0273 &   4 &  0.74 & $<\,$9.07  &  $<-$2.00  &  ...  \\
 29090 & J143810.20+092009.7 & 0.0303 &   4 &  0.62 & $<\,$9.08  &  $<-$1.93  &   *   \\
  7581 & J143915.52+024340.9 & 0.0279 &   4 &  0.70 & $<\,$9.06  &  $<-$1.98  &  ...  \\
  9748 & J143917.94+032206.0 & 0.0279 &   5 &  0.68 & $<\,$9.05  &  $<-$1.78  &  ...  \\
  9917 & J144025.99+033556.0 & 0.0281 &  10 &  0.46 & $<\,$8.88  &  $<-$1.78  &  ...  \\
  9704 & J144059.30+030813.5 & 0.0265 &   4 &  0.81 & $<\,$9.08  &  $<-$1.87  &  ...  \\
  9891 & J144225.70+031354.9 & 0.0258 &   4 &  0.74 & $<\,$9.02  &  $<-$1.95  &   *   \\
 29420 & J144858.71+122924.4 & 0.0474 &   5 &  0.66 & $<\,$9.51  &  $<-$1.75  &   *   \\
 10031 & J145106.43+045032.6 & 0.0273 &   8 &  0.55 & $<\,$8.94  &  $<-$1.89  &  ...  \\
 10040 & J145235.22+043245.1 & 0.0286 &   4 &  0.66 & $<\,$9.06  &  $<-$1.92  &  ...  \\
 10150 & J150026.75+041044.4 & 0.0328 &  34 &  0.24 & $<\,$8.73  &  $<-$1.80  &   *   \\
 10292 & J151322.09+040701.2 & 0.0426 &  90 &  0.15 & $<\,$8.77  &  $<-$1.67  &  ...  \\
 42020 & J151516.46+063918.5 & 0.0352 &  53 &  0.19 & $<\,$8.71  &  $<-$1.80  &   *   \\
 42017 & J151524.84+062654.0 & 0.0452 &  90 &  0.15 & $<\,$8.84  &  $<-$1.71  &  ...  \\
 42140 & J151531.97+072829.0 & 0.0457 &  18 &  0.32 & $<\,$9.16  &  $<-$1.82  &  ...  \\
 10367 & J151553.85+030301.1 & 0.0379 &   5 &  0.67 & $<\,$9.32  &  $<-$1.76  &   *   \\
 10358 & J151711.15+032105.7 & 0.0370 &   4 &  0.78 & $<\,$9.36  &  $<-$1.73  &   *   \\
 10404 & J151722.96+041248.9 & 0.0361 &   5 &  0.70 & $<\,$9.29  &  $<-$1.73  &   *   \\
 42156 & J151724.33+072921.9 & 0.0336 &  40 &  0.20 & $<\,$8.67  &  $<-$1.85  &   *   \\
 41974 & J151758.30+064445.4 & 0.0349 &  28 &  0.24 & $<\,$8.79  &  $<-$1.85  &   *   \\
 42175 & J151832.42+070720.7 & 0.0456 &  30 &  0.28 & $<\,$9.10  &  $<-$1.78  &  ...  \\
 10447 & J151840.93+042505.3 & 0.0471 &  87 &  0.16 & $<\,$8.88  &  $<-$1.80  &  ...  \\
 39469 & J151903.38+080819.4 & 0.0338 &   4 &  0.84 & $<\,$9.31  &  $<-$1.81  &  ...  \\
 39467 & J151953.53+080557.2 & 0.0334 &   5 &  0.66 & $<\,$9.20  &  $<-$1.78  &  ...  \\
 39465 & J152028.70+081706.6 & 0.0372 &  70 &  0.18 & $<\,$8.73  &  $<-$1.77  &   *   \\
 39448 & J152037.22+080305.7 & 0.0338 &  89 &  0.16 & $<\,$8.59  &  $<-$1.49  &   *   \\
 39605 & J152559.84+094724.5 & 0.0339 &  89 &  0.15 & $<\,$8.56  &  $<-$1.57  &  ...  \\
 39607 & J152706.28+094746.3 & 0.0438 &  20 &  0.32 & $<\,$9.13  &  $<-$1.77  &   *   \\
 39606 & J152716.73+094603.8 & 0.0437 &  69 &  0.17 & $<\,$8.84  &  $<-$1.81  &   *   \\
 26958 & J154654.33+055328.3 & 0.0419 &   4 &  0.81 & $<\,$9.49  &  $<-$1.77  &   *   \\
 21023 & J155636.91+272911.9 & 0.0415 &  89 &  0.16 & $<\,$8.78  &  $<-$1.54  &   *   \\
 10817 & J220120.93+121148.1 & 0.0291 &  15 &  0.38 & $<\,$8.83  &  $<-$1.78  &  ...  \\
 10872 & J221321.50+132611.3 & 0.0281 &  24 &  0.33 & $<\,$8.74  &  $<-$1.74  &  ...  \\
 10884 & J221430.63+130444.9 & 0.0257 &  18 &  0.31 & $<\,$8.63  &  $<-$1.84  &  ...  \\
\hline
\end{tabular}
\end{table*}

\begin{table*}
\centering
\caption{Weighted Average and Median Gas Fractions Plotted in Figure 9}
\label{t_avgs}
\begin{tabular}{lccccc}
\hline
 &  & $\langle M_{\rm HI}/M_\star \rangle$  & $\langle M_{\rm HI}/M_\star \rangle$   & \Mhi/\Mst &  \\
$x$  & $\langle x \rangle$ & (average)$^{a}$ & (average)$^{b}$ & (median)$^{c}$ & $\langle N \rangle^{d}$ \\
\hline
Log \Mst  	&    10.16 &   0.352$\pm$0.062 &   0.348$\pm$0.065 &   0.206 &   24   \\
 	  	&    10.49 &   0.124$\pm$0.026 &   0.117$\pm$0.027 &   0.049 &   37   \\
	  	&    10.77 &   0.078$\pm$0.023 &   0.072$\pm$0.024 &   0.030 &   51   \\
	  	&    11.05 &   0.050$\pm$0.011 &   0.042$\pm$0.011 &   0.017 &   60   \\
	  	&    11.29 &   0.024$\pm$0.011 &   0.014$\pm$0.013 &   0.016 &   15   \\
\hline
Log \must 	&     8.23 &   0.487$\pm$0.111 &   0.487$\pm$0.111 &   0.316 &   10   \\
	  	&     8.58 &   0.236$\pm$0.045 &   0.235$\pm$0.045 &   0.149 &   20   \\
	  	&     8.88 &   0.097$\pm$0.020 &   0.092$\pm$0.020 &   0.050 &   65   \\
 	  	&     9.15 &   0.036$\pm$0.007 &   0.026$\pm$0.007 &   0.017 &   73   \\
	  	&     9.36 &   0.020$\pm$0.006 &   0.011$\pm$0.008 &   0.017 &   12   \\
\hline
$R_{90}/R_{50}$  &     1.89 &   0.236$\pm$0.077 &   0.236$\pm$0.077 &   0.193 &   12   \\
		&     2.34 &   0.205$\pm$0.056 &   0.204$\pm$0.057 &   0.072 &   26   \\
 	 	&     2.70 &   0.154$\pm$0.024 &   0.149$\pm$0.024 &   0.070 &   52   \\
		&     3.10 &   0.059$\pm$0.011 &   0.049$\pm$0.012 &   0.018 &   77   \\
 		&     3.39 &   0.040$\pm$0.015 &   0.030$\pm$0.017 &   0.016 &   19   \\
\hline
\nuvr		&     2.89 &   0.280$\pm$0.036 &   0.280$\pm$0.036 &   0.208 &   27   \\
 		&     3.60 &   0.116$\pm$0.016 &   0.115$\pm$0.016 &   0.077 &   33   \\
 		&     4.36 &   0.093$\pm$0.021 &   0.090$\pm$0.022 &   0.048 &   29   \\
 		&     5.33 &   0.041$\pm$0.008 &   0.028$\pm$0.008 &   0.016 &   45   \\
 		&     5.86 &   0.019$\pm$0.002 &   0.006$\pm$0.003 &   0.016 &   41   \\
\hline
\end{tabular}
\begin{flushleft}
Notes. --- $^{a}$Weighted, average gas fraction; \hi\ mass of non-detections set to upper limit.
$^{b}$Weighted, average gas fraction; \hi\ mass of non-detections set to zero.
$^{c}$Weighted, median gas fraction; \hi\ mass of non-detections set to upper limit.
$^{d}$Average number of galaxies in bin.
\end{flushleft}
\end{table*}

\end{document}